\documentclass[prb,twocolumn,showpacs]{revtex4-1}

\usepackage{bm}
\usepackage{amssymb}
\usepackage{graphicx}
\usepackage{epstopdf}
\usepackage{amsmath}
\usepackage{color}
\usepackage{float}
\usepackage{braket}
\usepackage{comment}
\usepackage{epstopdf}
\usepackage{leftidx}

\newcommand{\BEQ}{\begin{equation}}
\newcommand{\EEQ}{\end{equation}}
\newcommand{\BEA}{\begin{eqnarray}}
\newcommand{\EEA}{\end{eqnarray}}
\newcommand{\Tr}{\mbox{Tr}}

\renewcommand{\a}{a}
\renewcommand{\b}{b}
\renewcommand{\c}{c}
\renewcommand{\d}{d}

\newcommand{\iu}{\mathrm{i}}
\newcommand{\eu}{\mathrm{e}}
\makeatletter
\newcommand\figcaption{\def\@captype{figure}\caption}
\makeatother

\newcommand{\Qp}{Q}
\newcommand{\Qm}{R}
\newcommand{\lsb}[1]{\leftidx{_{#1}}}

\renewcommand{\Re}{\text{Re}}
\renewcommand{\Im}{\text{Im}}

\begin{document}

\title{The Complex Spherical 2+4 Spin Glass: a Model for Nonlinear
  Optics in Random Media}

\author{F. Antenucci$^{1,2}$,
  A. Crisanti$^{1,3}$ and L. Leuzzi$^{2,1}$}
\email{luca.leuzzi@cnr.it} \affiliation{$^1$ Dipartimento di Fisica,
  Universit\`a di Roma ``Sapienza'', Piazzale A. Moro 2, I-00185,
  Roma, Italy\\ $^2$ IPCF-CNR, UOS {\it Kerberos} Roma, Piazzale
  A. Moro 2, I-00185, Roma, Italy \\ $^3$ ISC-CNR, UOS {\it Sapienza},
  Piazzale A. Moro 2, I-00185, Roma, Italy }

\begin{abstract}
A disordered mean field model for multimode laser in open and
irregular cavities is proposed and discussed within the replica
analysis.  The model includes the dynamics of the mode intensity and
accounts also for the possible presence of a linear coupling between
the modes, due, e.g., to the leakages from an open cavity.  The
complete phase diagram, in terms of disorder strength, source pumping
and non-linearity, consists of four different optical regimes:
incoherent fluorescence, standard mode locking, random lasing and the
novel spontaneous phase locking.  A replica symmetry breaking phase
transition is predicted at the random lasing threshold.  For a high
enough strength of non-linearity, a whole region with nonvanishing
complexity anticipates the transition, and the light modes in the
disordered medium display typical discontinuous glassy behavior,
i.~e., the photonic glass has a multitude of metastable states that
corresponds to different mode-locking processes in random lasers.  The
lasing regime is still present for very open cavities, though the
transition becomes continuous at the lasing threshold.
\end{abstract}


\maketitle

Considering salient multimode laser theory properties, both in ordered
and disordered amplifying media, in this paper we construct a general
statistical mechanical model for interacting waves. We study the
model, in particular, in the framework of nonlinear optics, focusing
on applications to the random lasing phenomenon.  The term ``random
lasing'' embraces a number of phenomena related to light amplification
by stimulated emission in systems characterized by a spatial
distribution of the electromagnetic field which is much more irregular
and complicated than for well-defined cavity modes of standard lasing
structures.  In any system, to produce a laser two ingredients are
essential: \emph{optical amplification} and \emph{feedback}.
Amplified spontaneous emission 
can occur even without
an optical cavity, and then the spectrum is determined only by the
gain curve of the active material.  Historically, already in the late
60's Letokhov \cite{letokhov1968generation} theoretically discussed
how light diffusion with gain can lead to the divergence of the
intensity above a critical volume, and, if the gain depends on the
wavelength, the emission spectrum narrows down close to the wavelength
of maximum gain.  These features were later observed in
experiments.\cite{Markushev86, gouedard1993generation} When the
multiple-scattering feedback dominates, instead, lasing occurs: 
a phenomenon known as \emph{Random Laser}
(RL).\cite{RL_def_96} The presence of feedback is associated with the
existence of well-defined long-lived cavity modes and characterized by
a definite spatial pattern of the electromagnetic field.  A RL is, in
other words, \emph{``mirror-less''} but not \emph{``mode-less''}
\cite{wiersma2008physics}.

Among the most singular aspects of RLs is that, for systems composed
by a large number of modes, a complex behavior in its temporal and
spectral response is observed: if there is no specific frequency that
dominates the others, the spectral resonances can change frequency
from one excitation pulse to another, with emission spectra strongly 
fluctuating from shot to shot  \cite{STS_Mujumdar_07, cit41_review,
  STS_vanDerMolen_06} and whose \emph{shot-to-shot} correlations 
appear to be highly non-trivial \cite{Ghofraniha15}.
 In these systems the scattering particle positions
and all external conditions are kept perfectly constant, so that these
differences can only be due to the spontaneous emission occurring when
the RL is activated at each pumping shot.
In these conditions it is observed that the intensity distribution is
not Gaussian, but rather of Levy type.\cite{cit44_review,STS_Lepri_07}
This is true close to the lasing threshold (where the mentioned
spectrum of fluctuations are expected \cite{wiersma2008physics}),
whereas far below and far above the threshold the statistics remains
Gaussian.

In the following we provide a general framework for the study of
physical systems described by complex amplitudes coupled by both
linear and nonlinear ordered or quenched disordered interaction terms.
We follow a statistical mechanical approach to derive general results
about the presence of a coherent regime and the type of transitions
involved.
Although the theory has a wide range of applications in modern
physics, e. g., to the Bose - Einstein condensation
\cite{PhysRevB.40.546}, we focus on non-linear optics and the random
lasing phenomenon.

The use of statistical mechanics of disordered systems for random
lasers was initially introduced in
Ref. [\onlinecite{AngelaniZamponiPRL}] for a phase model where, using
the \emph{replica method},\cite{MPVBook} was found that the
competition for the available gain of a large number of random modes
can lead to a behavior similar to that of a discontinuous glass
transition.\cite{Kirkpatrick87,Crisanti92,Goetze92,Leuzzi07}
Successive applications of the replica method to this
problem\cite{AngelaniZamponiPRB,ContiLeuzziPRL09,ContiLeuzziPRB11}
have shown that non-linear optics and random lasers can be a benchmark
for the modern theory of glassy systems.

In the present study we consider a more general and realistic model
for non-linear waves by removing the two basic assumptions of earlier
works: the quenched amplitude approximation and the strong cavity
limit.  We will consider the whole complex amplitudes of the
electromagnetic field eigenmodes expansion, not only the phases, as
the dynamic variables of the problem and we will account for the
presence of nonzero off-diagonal elements in the linear coupling, as
expected in presence of an irregular and/or open resonator.  The
inclusion of the intensity dynamics, in particular, could open new and
more practical ways to directly compare with experiments, cf., e.g.,
Ref. [\onlinecite{Ghofraniha15, Antenucci15}], since the mode intensities are
usually more easily accessible than the phases.  Some results of this
study have been presented in Ref. [\onlinecite{Antenucci14}].

The statistical mechanics of disordered systems provides a peculiar
point of view on this RL phenomenon: the leading mechanism for the
non-deterministic activation of the modes in this complex coherent
wave regime is identified in the frustration of the disordered
interactions and the consequent presence of many degenerate equivalent
glassy states.\cite{ContiLeuzziPRL09, ContiLeuzziPRB11}

The paper is organized as follows: in
Sect. \ref{sec:derivation_of_the_model} we present and discuss the
multimode laser theory for open resonators in the semiclassical limit
\cite{HackenbroichSemiclassical, ZaitsevDiagrammatic} that will be the
starting point for our statistical mechanical model.  In
Sect. \ref{sec:leading_model} we discuss the disordered mean field
model and define the control parameters.  In Sect. \ref{sec:results}
we report the results of the replica analysis of the disordered 
 model and discuss the type of phases and transitions predicted.
In particular, in Sect. \ref{sec:complexity} we examine the presence
of a region in the phase diagram with a nonzero complexity.  Finally
in Sect. \ref{sec:conclusions} we draw our conclusions.

\section{Multimode Laser Theory}
\label{sec:derivation_of_the_model}

The complex structure and the extreme openness of Random Lasers make these optical systems different from traditional cavity lasers. 
From a theoretical point of view, the strong coupling to the external world requires a different treatment from the standard approach of traditional laser textbooks.

The problem of describing quantum systems strongly interacting with the environment has large interest and it is relevant not only for the physics of lasers 
(see, e. g., Ref [\onlinecite{RotterReview}]).
The difficulty originates from the non-Hermiticity of the problem as the openness becomes relevant, so that the standard methods 
to solve or quantize Hermitian operators do not apply in this case.
The quantum 
system is localized in space. However, there is always a natural environment into which the quantum system with discrete states is embedded.
The environment consists of the continuum of extended scattering states into which the 
discrete states of the system are embedded and can decay.
The coupling matrix  between the discrete states of the system and the scattering 
states of the continuum determine the lifetime of the states, which is, therefore, usually finite.

Several approaches are presented in literature to build a set of modes suitable for a separation of time and coordinates dependencies of various physical observables,
in particular the electric and magnetic fields. \cite{FoxLi,QuasimodeDutra,TureciPRA06,ZaitsevReview}
Here, we start from the system-and-bath approach of Ref. [\onlinecite{HackenbroichViviescasPRA03}], 
in which a rigorous quantization of the field is possible.
We note, in particular, that the quantum treatment is necessary to compute the linewidth or the photon statistics of the output radiation.
In this approach, the contributions of radiative and localized
modes can be separated by the Feshbach  projection method onto two orthogonal subspaces.\cite{Feshbach}
This leads to an effective theory in the subspace of localized modes with  an effective 
linear off-diagonal damping coupling.\cite{HackenbroichViviescasPRL02,HackenbroichViviescasPRA03,HackenbroichViviescasJOB04}


The atom-field system can, then, be described
via the complex amplitudes of the localized electromagnetic modes $\alpha_\lambda$
and the atomic raising operator  $\sigma^\dagger_{-} =\ket{e} \bra{g} $ and inversion operator $\sigma_z = \ket{e} \bra{e} - \ket{g} \bra{g}$,
being $\ket{g}$ and $\ket{e}$ the ground and excited atom states.
%
The evolution of the operators can be expressed by the Jaynes-Cumming 
Hamiltonian,\cite{JC_Hamiltonian1,JC_Hamiltonian2}
and, including the cavity loss, is expressed in the Heisenberg representation as \cite{HackenbroichSemiclassical}
\begin{eqnarray}
  \dot{\alpha}_\lambda &=&
 - \iu \omega_\lambda 	\alpha_\lambda 
 -  \sum_\mu \gamma_{\lambda \mu} 
 \alpha_\mu
\label{eq:JK_1}
  \\
 &&
 + \int d \mathbf{r} \, g^\dagger_\lambda (\mathbf{r}) \sigma_- (\mathbf{r}) + F_\lambda \, , 
 \nonumber
 \\
  \dot{\sigma}_- (\mathbf{r}) &=&
 -( \gamma_\perp + \iu \omega_a ) \sigma_- (\mathbf{r}) 
 \label{eq:JK_2}
\\
 \nonumber
&& + 2 \sum_\mu g_\mu (\mathbf{r}) \sigma_z (\mathbf{r}) \alpha_\mu + F_- (\mathbf{r}) \, ,
  \\
  \dot{\sigma}_z(\mathbf{r}) &=&
 \gamma_\parallel \left( S \rho(\mathbf{r}) -\sigma_z(\mathbf{r}) \right)
 \label{eq:JK_3}
 \\
 &&
 -\sum_\mu \left( g_\mu^\dagger (\mathbf{r}) \alpha^\dagger_\mu \sigma_- (\mathbf{r})  + \text{h.c.} \right)   + F_z(\mathbf{r}) \, ,
\nonumber 
 \end{eqnarray}
where $\gamma_{\lambda \mu}$ is the damping matrix associated to the openness of the cavity, \cite{HackenbroichViviescasPRL02,HackenbroichViviescasPRA03}
$\rho(\mathbf{r})$ the atom density, 
$\omega_a$ the frequency of the atomic transition,
$\gamma_\perp$ the polarization decay rate,
 $\gamma_\parallel$ the population-inversion decay rate
and
 $S$ the pump intensity resulting from the interaction between the atoms and the external bath.
 The noise term $F_\lambda$ follows from the coupling with the bath.
 The field-atoms coupling constants are 
\begin{align}
 g_\lambda (\mathbf{r}) \equiv \frac{\omega_a p}{\sqrt{2 \hbar \epsilon_0 \omega_\lambda}}  \mu_\lambda (\mathbf{r}) \, ,
\label{eq:couplings_g_definition}
\end{align}
where $p$ is the atomic dipole matrix element and the $\mu_\lambda(\mathbf{r})$ are the orthogonal set of the resonator eigenstates. \cite{HackenbroichViviescasPRL02,HackenbroichViviescasPRA03}
The interaction also gives rise to the noises $F_- (\mathbf{r})$ and $F_z (\mathbf{r})$, due, for example, to the finite lifetime of the excited states for the decay to states not involved in the stimulated emission process.

The semiclassical theory consists in replacing the operators with their expectation values.
It is assumed that the lifetimes of the modes are much longer than the characteristic times of pump and loss:
the atomic variables can then be adiabatically removed to obtain the (non-linear) equations for the field alone.

Consider first the case of weak pumping, so that it is possible to assume $\sigma_z (\mathbf{r}) = S \rho (\mathbf{r}) $ and
the unique stationary solution is $\alpha_\lambda = 0$ for all the modes.
In this case the deviations from the stationary state relax to zero with complex frequency $\omega_k$
given by the eigenvalues of the non-Hermitian matrix \cite{HackenbroichSemiclassical}
\begin{align}
 & H_{\lambda \mu} =
 \omega_\lambda \delta_{\lambda \mu} 
 - \iu \gamma_{\lambda \mu}
 + \iu G^{(2)}_{\lambda \mu} (\omega_k) 
 \, ,
  \label{eq:G_linear_regime}
  \\
 &\text{with}
 \quad
 G_{\lambda \mu}^{(2)} (\omega) \equiv 2 S \int d \mathbf{r} \, \rho(\mathbf{r}) 
 \frac{g_\mu^* (\mathbf{r}) g_\lambda (\mathbf{r})}{\iu (\omega_a - \omega) + \gamma_\perp} \, .
 \nonumber
\end{align}
In general, if the cavity is open and/or the atoms are not uniformly distributed in the resonator, the matrix $H_{\lambda \mu}$ 
is not diagonal:
the eigenvalues and eigenvectors are, hence, different from the case of cold cavity and depend parametrically on the pump strength $S$.
In particular, increasing $S$ the eigenvalues move up in the complex plane. The lasing threshold is reached when
one eigenvalue takes a positive imaginary part. In this case the gain exceeds the loss and the solution 
$\alpha_\lambda = 0$ becomes unstable.

At the lasing threshold is, then, necessary to consider the time evolution of the atom operators that provides an effective non-linear coupling among the electromagnetic modes.
In this case the standard approach is to consider an expansion in power of the mode amplitudes.
One starts neglecting the quadratic term in Eq. (\ref{eq:JK_3}) obtaining the zero-order approximation, that replaced in Eq. (\ref{eq:JK_2}) gives the first order approximation
that replaced back in Eq. (\ref{eq:JK_3}) gives the second-order approximation and so forth.
Inserting the result into Eq. (\ref{eq:JK_1}) one  can construct a perturbative approximation of the effective evolution of the mode amplitudes.

In the particular case of the \emph{free-running approximation}, \cite{HackenbroichSemiclassical}
that is assuming that the different lasing modes oscillate independently from each other (so that the phases are uncorrelated and the interaction concerns the intensities alone), 
it is possible to resum the equation and obtain an expression for the mode intensities valid to all the orders in the perturbation theory (cf. Ref. [\onlinecite{ZaitsevDiagrammatic}]).
This approximation may be valid for the so-called non-resonant or incoherent feedback emission in disordered cavities,\cite{letokhov1968generation} where the interference effects do not play any role.
In this case the emission is due solely to amplified spontaneous emission, and, then, the spectrum is determined only by the gain curve of the active material.
This  approach explains some simple properties of emission from disordered cavities. \cite{Markushev86, gouedard1993generation}
For the lasing regime, however, it is the multiple-scattering induced feedback that defines optical modes, with a well resolved frequency, a given bandwidth and spatial 
profile.\cite{wiersma2008physics}
 It, thus, becomes essential to include the phases into the analysis and consider the non-linear interactions  non-perturbatively.

Since we are mainly interested in the characterization of the random lasing regime, 
we do not assume the free-running approximation and limit ourselves to the non-linear third order theory. The subsequent orders may become relevant far above the threshold. From a statistical mechanics point of view 
the orders beyond the third are not expected to change the universality class of the transition
for a large class of models (see, e.~g., Ref. [\onlinecite{CrisantiLeuzzi_sp_NPB}]). 
Being specific, if one  considers 
$g^2 |a|^2 \ll  \gamma_\perp \gamma_\parallel$, where $|a|^2$ is the typical intensity in the lasing regime, 
the third order theory is exact.

\subsection{Cold Cavity vs Slow Amplitude Modes}

The evolution in the lasing regime is conveniently expressed in the basis of the slow amplitude modes.
A \emph{slow amplitude mode} with index $l$ is a solution such that it has a harmonic form for $t \gg 1$ and, therefore, its Fourier transform is proportional to $\delta (\omega-\omega_l)$.
By definition, a \emph{lasing mode} is a slow amplitude mode with a positive intensity at the solution.
In general the lasing modes are different from the cold cavity ones. 
The steady-state solutions are different already in the linear regime, as $G^{(2)}_{\lambda \mu}$ is not diagonal, cf. Eq. (\ref{eq:G_linear_regime}). 
We express the relationship between cold cavity modes $\alpha_\lambda$  and laser modes $a_k$  in the form
\begin{equation}
  \alpha_\lambda (t) = \sum_k A_{\lambda k} \overline{a}_k (t) 
  ,  \qquad
 \overline{a}_k (t) = a_k (t) \, \eu^{- \iu \omega_k t} 
 \label{eq:lasing_modes_decomposition}
\end{equation}
with $a_k (t)$ evolving on time scales much longer than $\omega_k^{-1}$, so that
$ \overline{a}_k (\omega) \simeq  \delta (\omega - \omega_k) $.
Here and in the following we use Greek letters for cold cavity modes and Latin letters for the slow amplitude modes.

We consider a complete \emph{slow amplitude modes basis} in order to expand any mode and, in particular, invert  Eq. (\ref{eq:lasing_modes_decomposition})
\begin{align}
 & \overline{a}_k (t) = \sum_\lambda B_{k \lambda}^* \, \alpha_\lambda (t) 
  \quad , &
 &  B_{k \lambda}^* 
 =
 \left( A_{k \lambda} \right)^{-1} \, .
 \label{eq:decomposition_inverse}
\end{align}
Using the expansion in the slow amplitude modes 
we can express the mode evolution Eq. (\ref{eq:JK_1}) at the third order as
\begin{eqnarray}
 \label{eq:langevin_equation_lasing_modes}
  \dot{a}_l(t) &= &
 \sum_{k \mid \text{FMC} (l,k)} \Bigl[ 
\tilde{\gamma}_{lk} - S\, G^{(2)}_{lk}
 \Bigr] a_k (t)
 \\
 &&
 \nonumber - S 
  \sum_ { \mathbf{k}  \mid  \text{FMC} (l,\mathbf{k})  }  
  G_{l \mathbf{k}}^{(4)}\,
  a_{k_1} (t)\,
 a_{k_2}^* (t)\,  
 a_{k_3} (t) 
 \, ,
\end{eqnarray}
where the sums are restricted to terms that meet
the \emph{frequency matching conditions} (FMC) (see below),
the matrix $\tilde{\gamma}_{lk} $
is 
\begin{align}
 \tilde{\gamma}_{lk} &\equiv \sum_{\lambda \mu} B^*_{\lambda l} \gamma_{\lambda \mu} A_{\mu k}
 \, ,
\label{eq:damping_laser}
\end{align}
with the left and right coupling constants for the slow amplitude modes  given by
\begin{align}
 &g^{L }_{k} = \sum_\mu B_{\mu k} g_\mu
 \, , &
 &g^{R }_{k} = \sum_\mu A_{\mu k} g_\mu \, ,
\end{align}
and  $G_{lk}^{(2)}$ and $G_{l\mathbf{k}}^{(4)}$ are functions of the frequencies 
$\omega_{k}$:
\begin{eqnarray}
\label{eq:G2_coldcavity}
  G_{l k}^{(2)} =&  M^{(2)}_k \int d \mathbf{r} \, \rho(\mathbf{r}) \, g_l^{L *} (\mathbf{r}) \, g_k^{R} (\mathbf{r})
 \, , 
 \\
 G_{ l \mathbf{k}}^{(4)} = &
M^{(4)}_{\mathbf{k}}
 \int d \mathbf{r} \rho(\mathbf{r}) 
 g^{L *}_l (\mathbf{r})  g_{k_1}^R (\mathbf{r}) g_{k_2}^{R *} (\mathbf{r})  g_{k_3}^R (\mathbf{r})  \, .
 \label{eq:G4_coldcavity}
\end{eqnarray} 
The coefficients $  M^{(2)}_{ k}$ and $  M^{(4)}_{ \mathbf{k}}$ are defined as
\begin{eqnarray}
 M_k^{(2)} &\equiv &
 - \frac{1 }{ \pi \gamma_\perp }
  \, D(\omega_k) \, ,
 \nonumber
 \\
\label{eq:definition_nonlinearCoupling}
  M^{(4)}_{ \mathbf{k}}
  &\equiv &
   \frac{D(\omega_{k_3}) + D^* (\omega_{k_2})  }{2 \pi^3 \gamma_\perp^2 \gamma_\parallel }
   \\
   &&
   \nonumber
\times   D(\omega_{k_1}-\omega_{k_2}+\omega_{k_3}) 
   D_\parallel (\omega_{k_3}-\omega_{k_2}) 
  \end{eqnarray}
with
\begin{align}
  & D_\parallel (\delta\omega) \equiv \left(  1 - \iu \frac{\delta\omega}{\gamma_\parallel}  \right)^{-1} 
  , &
 D(\omega) \equiv \left( 1- \iu \frac{\omega-\omega_a}{\gamma_\perp} \right)^{-1} 
 \label{eq:D}
\end{align}
The notation 
$\sum_{\mathbf{k} \mid\text{FMC} (l,\mathbf{k})}$ introduced  in  
Eq. (\ref{eq:langevin_equation_lasing_modes})  denotes that  
the  slow amplitude condition $\overline{a}_k (\omega) \simeq
\delta(\omega-\omega_k)$ restricts the sums to modes $(l,\mathbf{k})$ that meet the 
the frequency matching condition. For generic $k_1\dotsc,k_{2n}$ interacting modes
the FMC reads:
\begin{align}
\label{eq:frequencyMatching}
 \text{FMC}(\mathbf{k}) \ : \
| \omega_{k_1} - \omega_{k_2} + \ldots + \omega_{k_{2n-1}} - \omega_{k_{2n}} | \lesssim \gamma \, .
\end{align}
The finite linewidth $\gamma$ of the modes can be thoroughly derived only in a complete quantum theory.
In particular, the noise factors in Eqs. (\ref{eq:JK_1})-(\ref{eq:JK_3}) have to be included,
resulting in a weak time dependence of $\overline{a}_k(t)$ in Eq. (\ref{eq:lasing_modes_decomposition}).
Here, we include it in an effective way, as a parameter to suitably conform to different experimental  situations.

We stress as, in general, the linear term of Eq.
(\ref{eq:langevin_equation_lasing_modes}) may have non-zero
off-diagonal terms.  They are all zero when the frequencies are well
distinct, i.e., the spectral interspacing $\delta \omega \gg \gamma$,
so that the frequency matching condition of the linear term is never
satisfied but for the modes with overlapping frequency.  While this is
generally true for standard high quality-factor lasers,
\cite{HausPaper} for RL there can be a significant frequency overlap
between the lasing modes, $\delta \omega \sim \gamma$, and off-diagonal
linear contributions must be considered in the slow amplitude basis.

The actual values of the couplings are in principle, and in some simple case, entirely computable in
the cold cavity basis, cf. Eqs. (\ref{eq:G2_coldcavity})-(\ref{eq:G4_coldcavity}) and, e.~g., Ref. [\onlinecite{HackenbroichViviescasJOB04}]. 
The main problem remains how
to express the interactions in the slow amplitude mode basis actually used in the dynamics. 
In some cases the solution can be found using some self-consistent procedures proceeding 
iteratively starting from the solution
obtained without the non-linear coupling.\cite{Tureci08, TureciAbInitio09, Rotter14}
In particular, when the non-linear term is entirely neglected, a possible (though not unique) solution is the one that
diagonalizes the linear interaction.
Nonetheless, when the lasing threshold is exceeded, the non-linear  term becomes non-perturbatively relevant
and the diagonalization of the linear term does not correspond to a slow amplitude basis 
in the most general case of lasing in random media.

\subsection{The role of the noise }
\label{sec:hamiltonian_formulation}
In the previous semiclassical derivation we have neglected all noise
sources.  However, to obtain a complete statistical description the
noise, and, hence, the spontaneous emission and the heat-bath
temperature, must be taken into account.  Indeed, we will show that
the role of the entropy becomes crucial for disordered multimode
lasers, where a random first order transition \cite{RFOT} is expected,
at least in the mean-field approximation.

In general, different noise sources occur (see $F_\lambda$, $F_-$ and
$F_z$ in Eqs. (\ref{eq:JK_1})-(\ref{eq:JK_3})).  Further on, in the
case of open cavities, it is known that the noise $F_\lambda$ due to
the external bath coupling is correlated in the cold cavity modes
basis. \cite{HackenbroichViviescasPRL02, HackenbroichViviescasPRA03}
We, then, consider the presence of a noise $F_l(t)$ in
Eq. (\ref{eq:langevin_equation_lasing_modes}):
\begin{align}
\nonumber
 & \langle F_{k_1}^* (t_1) \, F_{k_2} (t_2) \rangle = 2 T \, \Gamma_{k_1 k_2 } \, \delta (t_1-t_2) \, ,
 \\
 & \langle F_{k_1} (t_1) \, F_{k_2} (t_2) \rangle = 0 \, ,
 \label{eq:correlated_noise}
\end{align}
with $T$ being the spectral power of the noise, proportional to the
heat-bath temperature.
The matrix  $\Gamma_{k_1 k_2}$ 
corresponds to the damping matrix
Eq. (\ref{eq:damping_laser}) for $F_\lambda$ and, in general, is non
diagonal for open cavities.  
It can be, though, diagonalized, at the possible
price of having a non-diagonal linear contribution in
Eq. (\ref{eq:langevin_equation_lasing_modes}).  
Indeed, a non-unitary change of basis affects the noise
correlation:
\begin{align}
 a_l =& \sum_\lambda A^{-1}_{l \lambda} \alpha_\lambda 
\quad \to \quad 
F_l = \sum_\lambda A^{-1}_{l \lambda} F_\lambda 
\nonumber
\end{align}
and
\begin{eqnarray}
\langle F_{k_1}^* (t_1)  F_{k_2} (t_2) \rangle
=
\sum_{\lambda_1 \lambda_2}
\left( A^{*} \right)^{-1}_{l_1 \lambda_1} 
\langle F_{\lambda_1}^* (t_1) \, F_{\lambda_2} (t_2) \rangle 
A^{-1 }_{l_2 \lambda_2} 
\nonumber
\end{eqnarray}
The decomposition in the slow amplitude modes,
Eq. (\ref{eq:lasing_modes_decomposition}), is by no means unique.
This freedom may be used to build a mode basis where the noise
is  uncorrelated.
In the following, we assume that the various independent noise sources
act so that such basis construction is possible and, hence, the noise
can be assumed white and uncorrelated also in the general case of open
and irregular cavities.
We will, consequently, 
consider Gaussian, white and uncorrelated noise:
\begin{align}
\nonumber
 & \langle F_{k_1}^* (t_1) \, F_{k_2} (t_2) \rangle = 2 T \, \delta_{k_1 k_2 } \, \delta (t_1-t_2) \, ,
 \\
 & \langle F_{k_1} (t_1) \, F_{k_2} (t_2) \rangle = 0 \, ,
 \label{eq:uncorrelated_noise}
\end{align}
However, since diagonalization of the noise terms may 
result in possible off-diagonal terms in
the linear coupling in Eq. (\ref{eq:langevin_equation_lasing_modes}),
independent from the presence of the non-linear interaction, 
off-diagonal
linear interaction is, thus, considered throughout the rest of the paper.

In analogy with the standard mode locking case, \cite{GordonFisherPRL}
we, hence, eventually define the complex valued functional
\begin{align}
 \nonumber
 \mathcal{H}
 =&
 - \sum_{\mathbf{k}  \mid  \text{FMC} (\mathbf{k}) } \left[ 
 \tilde{\gamma}_{k_1 k_2} 
 -  S\, G^{(2)}_{k_1 k_2}
 \right] \, a_{k_1}^* a_{k_2}
\nonumber
\\
& +  \sum_ { \mathbf{k}  \mid  \text{FMC} (\mathbf{k})  }  
 S\, G^{(4)}_{k_1 k_2 k_3 k_4}  \,  
   a_{k_1} \,
   a_{k_2}^* \,
   a_{k_3} \,
   a_{k_4}^* 
   \nonumber
 \\
  \equiv &  
   \sum_{\mathbf{k}   \mid  \text{FMC} (\mathbf{k}) } g^{(2)}_{k_1 k_2} \, a_{k_1} a^*_{k_2} 
   \nonumber 
   \\
  & +
  \frac{1}{2}  \sum_ { \mathbf{k}  \mid  \text{FMC} (\mathbf{k})  }  
  g^{(4)}_{k_1 k_2 k_2 k_4}\,  a_{k_1} a_{k_2}^* a_{k_3} a_{k_4}^* 
\label{eq:general_hamiltonian}
\end{align}
where again the sums are restricted by the FMC \eqref{eq:frequencyMatching},
yielding the complex Langevin equation for the stochastic dynamics of
the amplitudes
\begin{equation}
\dot a_k = -\frac{\partial {\cal H}}{\partial a_k^*}+F_l
\label{eq:cLang}
\end{equation}
Eq. (\ref{eq:general_hamiltonian}) can
be rewritten in terms of its real and imaginary parts ${\cal H}_R+\iu
{\cal H}_I$:
\begin{align*}
 \mathcal{H}_R
 =& 
   \sum_{\mathbf{k}   \mid  \text{FMC} (\mathbf{k})}  G_{k_1 k_2}  \, a_{k_1} a^*_{k_2} 
   \\
   &
   \nonumber
   +
  \frac{1}{2} \sum_ { \mathbf{k}  \mid  \text{FMC} (\mathbf{k})  }  
   \Gamma_{k_1 k_2 k_3 k_4} \, a_{k_1} a_{k_2}^* a_{k_3} a_{k_4}^* 
  \, ,
 \\
  \mathcal{H}_I
 =& 
   \sum_{\mathbf{k}   \mid  \text{FMC} (\mathbf{k})} D_{k_1 k_2}  \, a_{k_1} a^*_{k_2} 
   \\
   &\nonumber+
  \frac{1}{2} \sum_ { \mathbf{k}  \mid  \text{FMC} (\mathbf{k})  }  
  \Delta_{k_1 k_2 k_3 k_4}  \, a_{k_1} a_{k_2}^* a_{k_3} a_{k_4}^* 
  \, ,
\end{align*}
with
\begin{align}
\label{eq:def_G}
  G_{k_1 k_2}  \equiv &
 \frac{1}{2} \left( g^{(2)}_{k_1 k_2} + g^{(2) *}_{k_2 k_1} \right)
 \, , 
 \\
  \Gamma_{k_1 k_2 k_3 k_4}  \equiv &
 \frac{1}{2} \left( g^{(4)}_{k_1 k_2 k_2 k_4} + g^{(4) *}_{k_2 k_1 k_4 k_3} \right)
 \label{eq:def_Gamma}
 \\
  \iu D_{k_1 k_2} \equiv &
 \frac{1}{2} \left(  g^{(2)}_{k_1 k_2} - g^{(2) *}_{k_2 k_1} \right)
 \, , 
 \label{eq:def_D}
 \\
  \iu \Delta_{k_1 k_2 k_3 k_4} \equiv &
 \frac{1}{2} \left( g^{(4)}_{k_1 k_2 k_3 k_4} - g^{(4) *}_{k_2 k_1 k_4 k_3} \right)
 \, .
 \label{eq:def_Delta}
\end{align}
Considering, as well,
 real and imaginary parts of mode amplitudes $a_l \equiv \sigma_l + \iu \tau_l$, 
 the stochastic  Eq. (\ref{eq:cLang}) can  be expressed as
\begin{align}
 & \frac{\partial \sigma_l}{\partial t}
 = - \frac{1}{2} \frac{\partial \mathcal{H}_R}{\partial \sigma_l} 
   + \frac{1}{2} \frac{\partial \mathcal{H}_I}{\partial \tau_l} 
   + F^R_l
   \, ,
   \nonumber
   \\
 & \frac{\partial \tau_l}{\partial t}
 = - \frac{1}{2} \frac{\partial \mathcal{H}_R}{\partial \tau_l} 
   - \frac{1}{2} \frac{\partial \mathcal{H}_I}{\partial \sigma_l} 
   + F_l^I
   \, .
\label{eq:langevin_real_and_immaginary}
\end{align}
From Eqs. (\ref{eq:langevin_real_and_immaginary})
it is clear that $\mathcal{H}_R$ is associated with a purely dissipative motion (a gradient flow in the $2N$ dimensional space $\sigma_1 , \ldots \sigma_N ,  \tau_1, \ldots \tau_N$ ),
while $\mathcal{H}_I$ generates a purely Hamiltonian motion for the $N$ conjugated variables $(\sigma_l , \, \tau_l )$.
If $\mathcal{H}_R = 0$ the total optical intensity $\mathcal{E} \equiv \sum_k |a_k |^2$ is a constant of motion under the previous Langevin equations (like $\mathcal{H}$ itself).
When $\mathcal{H}_R \neq 0$ this is no longer true, though the system is still stable 
because the gain decreases as the optical intensity increases.\cite{Chen_94}
For standard lasers this is usually modeled assuming that the gain is such that 
\begin{equation}
G_{kk} = \frac{G_0 }{1+\mathcal{E}/E_{\text{sat}}}, \quad \forall k
\end{equation}
where $E_{\text{sat}}$ is the saturation power of the amplifier.
To study the equilibrium properties of the model, it is possible to consider a simpler model:
at any instant the gain is supposed to assume exactly the value that  keeps $\mathcal{E}$ a constant of the motion,
as Gordon and Fisher have proposed in Ref.  [\onlinecite{GordonFisherPRL}].
In this way the system evolves over the hypersphere $\mathcal{E} \equiv \mathcal{E}_0$.

The relation between the thermodynamics in the fixed-power ensemble and a variable-power ensemble
might be seen as similar to the relation between the canonical and grand canonical ensembles in statistical mechanics.\cite{GatGordonFisher04}
The constraint $\mathcal{E} \equiv \mathcal{E}_0$ will induce a correlation of order $N^{-1}$ in the noise $F_l$.
However, as far as we are interested in the limit $N \gg 1$, such correlation can be neglected and the noise considered as white.

The request $\partial \mathcal{E} / \partial t = 0 $
implies that $G_0$, expressed as $ G_{k k} \equiv G_0 + G^{ \delta}_{k k} $, is given by
\begin{align*}
 \mathcal{ E}G_0 = &
 -
   \sum_{\mathbf{k}  \mid  \text{FMC} (\mathbf{k})}  G^{ \delta}_{k_1 k_2} \, a_{k_1} a^*_{k_2} 
  \\
  &-
   \sum_ { \mathbf{k}  \mid  \text{FMC} (\mathbf{k})  }  
  \Gamma_{k_1 k_2 k_3 k_4} \, a_{k_1} a_{k_2}^* a_{k_3} a_{k_4}^*  
 \, .
\end{align*}
In particular for $G^{ \delta}_{k_1 k_2} = G^{ \delta}_{k_1} \delta_{k_1 k_2}$ and 
$\Gamma_{k_1 k_2 k_3 k_4} = \Gamma$, the known result 
for the standard mode-locking case is recovered.\cite{GordonFisherPRL}
Inserting this expression for $G_0$ in the Langevin equations  (\ref{eq:cLang}),
one finds that in this case the  functional $\mathcal{H}_R$   becomes  
\begin{align}
\label{eq:effective_Hamiltonian}
 \mathcal{H}_R \equiv &
 - \frac{\mathcal{E}_0}{\mathcal{E}} 
 \sum_{\mathbf{k}  \mid  \text{FMC} (\mathbf{k})}  G^{ \delta}_{k_1 k_2} \, a_{k_1} a^*_{k_2} 
 \\
 \nonumber
 &
 - \frac{\mathcal{E}_0^2}{2 \mathcal{E}^2}
 \sum_ { \mathbf{k}  \mid  \text{FMC} (\mathbf{k})  }  
  \Gamma_{k_1 k_2 k_3 k_4} \, a_{k_1} a_{k_2}^* a_{k_3} a_{k_4}^*  \, .
\end{align}
where now coupling coefficients $G^\delta$ and $\Gamma$  do not depend from complex amplitudes $a$'s.
This expression for the Hamiltonian includes the condition that
the total optical power $\mathcal{E}$ is a constant of motion.
Imposing the spherical constraint simplifies  the coefficients ${\cal E}_0/{\cal E}=1$.

\subsection{Purely Dissipative Case}
In the case $\mathcal{H}_R \gg \mathcal{H}_I$
the functional $\mathcal{H}$ is approximately real. 
For the standard mode-locking lasers this corresponds to the physical situation where
the group velocity dispersion and the Kerr effect can be neglected.\cite{HausPaper}
The purely dissipative case does also apply to the important case of soliton 
lasers.\cite{GordonFisherOC}
In the general case of Eq. (\ref{eq:general_hamiltonian}) the situation is more complex. 
If the coefficients $g_{k_1 k_2}^{(2)}$ and $g_{k_1 k_2 k_3 k_4}^{(4)}$ are real
then the imaginary part of ${\cal H}$ clearly vanishes, see Eqs. (\ref{eq:def_D},\ref{eq:def_Delta}).
The requirement of real coefficients is, however, not necessary to ensure  a real valued 
${\cal H}$, cf. Eq. (\ref{eq:general_hamiltonian}), of the form given by Eq. (\ref{eq:effective_Hamiltonian}).
A less strict, though sufficient condition is that, cf. Eqs. (\ref{eq:lasing_modes_decomposition}), (\ref{eq:D}), 
\begin{align}
& A_{\lambda l}=A^*_{l\lambda} 
 \, ; &
 & \omega_l - \omega_a \ll \gamma_\perp 
 \, ; &
 & \delta \omega \ll \gamma_\parallel
\end{align}
This is consistent with, but stronger than, the usual rotating-wave approximation, $\delta \omega \ll \omega_l$,  $\forall l$.

The case with a real functional $\mathcal{H}$ is of particular interest because it can be studied using the standard methods of the equilibrium statistical physics.
In fact, when the functional $\mathcal{H}$ is real, the Eqs. (\ref{eq:langevin_real_and_immaginary}) 
reduce to the familiar ``potential form'': 
the evolution is the derivative of a ``potential'' respect to the considered variables plus white Gaussian noise.
Hence, the steady-state solution of the associated Fokker-Plank equation
\begin{align}
 \dot{\rho} =& 
 - \sum_k  \frac{\partial}{\partial \sigma_k}  \left\{ \frac{\partial \mathcal{H}_R}{\partial \sigma_k}  \rho \right\}
 - \sum_k \frac{\partial}{\partial \tau_k} \left\{ \frac{\partial \mathcal{H}_R}{\partial \tau_k}  \rho \right\}
\nonumber
\\
& + T \sum_k \left( \frac{\partial^2}{\partial \sigma^2_k} + \frac{\partial^2}{\partial \tau^2_k} \right) \rho
\end{align}
is given by the familiar Gibbs distribution 
\begin{align}
 \rho \left(  \sigma_1 , \tau_1 \ldots  \sigma_N , \tau_N \right) = \frac{\eu^{- \mathcal{H}_R/T_{\rm ph}}}{\int \eu^{- \mathcal{H}_R/T_{\rm ph}} d \sigma_1 d \tau_1 \ldots d \sigma_N d \tau_N}  \, .
\label{eq:gibbs_distribution}
\end{align}
where $T_{\rm ph}$ is an effective "photonic'' temperature proportional to the heat-bath temperature.
This case, then, is the most interesting for the application of statistical mechanics and it will analyzed in this paper.

The general case of $\mathcal{H}_I \neq 0$ is harder to study analytically.
We note that, in general, it is expected that transitions of the first order are not removed by slight modification of the dynamics.
The  analysis of the complete complex Langevin dynamics is postponed to a future work.

\section{The Statistical Mechanics Approach}
\label{sec:leading_model}

In the rest of this work we shall discuss the properties of mean-field solution of the model described 
by the Hamiltonian Eq. (\ref{eq:effective_Hamiltonian}) with the global {\em
  spherical} ~\footnote{It is said spherical using the naming adopted in
  spin systems \cite{Berlin52} where spins $\sigma_i$ are approximated
  by continuous real fields taking values on the $N$-dimensional
  hypersphere $\sum_{i=1}^N\sigma_i^2$. Here, in the photonic
  amplitude model, unlike the classic spherical spin model, the global
  spherical constraint is formally in the $\mathbb{C}^N$ space of $N$
  complex degrees of freedom.}  
constraint $\mathcal{E} \equiv \sum_k |a_k|^2 = \mathcal{E}_0 $.
 We notice that, because of this  constraint on the power,
adding a constant diagonal term to the pairwise coupling is irrelevant
for the thermodynamics of the system.

The mean-field solution  is exact when the probability distribution of the couplings is the same for all the mode couples $(k_1 , \, k_2)$
and quadruplets $(k_1 , \, k_2 , \, k_3, \, k_4)$.
This corresponds to the physical situation in which the two following conditions hold:
\begin{itemize}
 \item \emph{narrow-bandwidth:}
 the linewidth of the mode frequencies $\gamma$ is comparable with the total emission bandwidth $\Delta \omega$ so that
 the frequency-matching conditions are always satisfied, cf. Eq. (\ref{eq:frequencyMatching});
 \item \emph{extended modes:}
 all the mode localizations are extended to a spatial region  which scales 
 with the total volume $V$ of the active medium.
\end{itemize}

The first condition also implies that the diagonal elements of $G_{k_1 k_2}$ are all equal
and do not depend on the frequency.
In particular, then, for a \emph{strong cavity} and a regular medium these disappear from the equilibrium dynamics because of the spherical constraint.\cite{AngelaniZamponiPRB, ContiLeuzziPRB11}

The couplings are related to the spatial overlaps of the modes, 
cf. Eqs. (\ref{eq:G2_coldcavity})-(\ref{eq:G4_coldcavity}), and are, therefore, not independent. 
However, in the mean-field limit  off-diagonal correlations vanish for large system sizes
and we can assume that  the couplings are statistically independent.

Before discussing the properties of this solution we first rewrite the Hamiltonian
in the mean-field form
\begin{equation}
\label{eq:effective_Hamiltonian_3}
 \mathcal{H} = - \frac{1}{2} \sum_{j k}^{1,N} J_{j k} a_{j} a^\ast_{k} 
 - \frac{1}{4!} \sum_{j k l m}^{1,N} J_{j k l m} a_{j} a_{k} a^\ast_{l} a^\ast_{m}, 
\end{equation}
where $a_i$ are $N$ complex amplitude variables subject to the spherical constraint
 $ \sum_k | a_k |^2 = \mathcal{E}_0 \equiv \epsilon N$.
The couplings $J_{i_1,\dotsc,i_p}$ ($p=2,4$) are symmetric under index permutation and 
vanish if two or more indexes are equal. The non-null  $J_{i_1,\dotsc,i_p}$
are quenched independent identically distributed real random variables.
In the mean-field limit only the first two moments are relevant, so we can take 
a Gaussian distribution:
\begin{equation}
\label{eq:disorder}
 \mathcal{P} \left( J_{i_1 \ldots i_p} \right) =
 \frac{1}{\sqrt{2 \pi \sigma_{p}^2} } \exp \left[ - \frac{\left( J_{i_1 \ldots i_p} - \tilde{J}_0^{(p)} \right)^2}{2 \sigma_{p}^2} \right] \, .
\end{equation}
The requirement of an extensive  Hamiltonian,  i.e., proportional to $N$, 
requires that the variance
and the average
scale with $N$ as
\begin{align*}
 \sigma_{p}^2 & = \frac{p! \, J^2_p}{2 N^{p-1}} \, , &
 \tilde{J}_0^{(p)} = & \frac{J_0^{(p)}}{N^{p-1}}\, ,
\end{align*}
where  $J_p$ and $J_0^{(p)}$ are intensive parameters fixing the relative strength of various terms in the 
Hamiltonian.
To have a direct interpretation in terms of optical quantities we express them as 
\begin{align}
 J_0^{(4)} =& \alpha_0 J_0 
 \, , &
 J_0^{(2)} =& (1-\alpha_0) J_0 \, ,
 \\
 J_4^2 =& \alpha^2 J^2
 \, , &
 J_2^2 =& (1-\alpha)^2 J^2 \, .
\end{align}
where the parameters $J_0$ and $J$ fix the cumulative strength of the ordered and disordered
part of the Hamiltonian, while $\alpha_0$ and $\alpha$ the degree (i. e., relative strength) of the non-linear quartic ($p=4$) contribution in the ordered and disordered parts, respectively.
Then we introduce the usual parameters of RL models: the \emph{degree of disorder} $R_J$ 
and the \emph{pumping rate} $\mathcal{P}$ as:\cite{ContiLeuzziPRB11}
\begin{align}
 R_J & \equiv \frac{J}{J_0}
 \, , &
 \mathcal{P} \equiv \epsilon \sqrt{\beta J_0},
\label{eq:def_pumping_cap3}
\end{align}
where $\beta = T^{-1}$ is the ordinary inverse thermal bath temperature, cf. Eq. (\ref{eq:uncorrelated_noise}).
The definition of the pumping rate encodes the experimental fact that the effect of decreasing the temperature of the bath or increasing 
the energy of the pump source is qualitatively the same on the onset of a random lasing regime. \cite{Wiersma2001,Nakamura10}
We stress that the effective ``photonic" temperature $T_{\rm ph}\equiv T/\epsilon^2$, coupled to ${\cal H}$ in the Gibbs measure
Eq. (\ref{eq:gibbs_distribution}) in units of $J_0$ is nothing else than $\sqrt{\cal P}$.

With this parametrization the mean-field solution is conveniently expressed through the 
parameters
\begin{align}
\nonumber
 b_2 &= \frac{1-\alpha_0}{4} \mathcal{P} \sqrt{\beta J_0} 
 \, , &
 b_4 &= \frac{\alpha_0}{96} \mathcal{P}^2 \, ,
 \\
 \xi_2 &= \frac{(1-\alpha)^2}{4} \beta J_0 \mathcal{P}^2  R_J^2  
 \, , &  
   \xi_4 &= \frac{\alpha^2}{6} \mathcal{P}^4  R_J^2,
\label{eq:standard_to_photonics}
\end{align}
which are the \emph{photonics}  counterpart of the standard $p$-spin-like \cite{Crisanti92}, or mode coupling theory like \cite{Goetze09} parameters 
\begin{align}
 & b_2 =  \frac{\epsilon}{4} \beta J_0^{(2)}   \; , &  
 & b_4 =  \frac{\epsilon^2}{96} \beta J_0^{(4)} \, ,
 \nonumber
 \\
 & \xi_2 =  \frac{\epsilon^2}{4} \beta^2 J_2^2  \;  , &  
 & \xi_4 =  \frac{\epsilon^4}{6} \beta^2 J_4^2   \, ,
 \label{eq:definition_parameters_b_xi_mu}
\end{align}
used in the statistical mechanics study of these type of models.
Notice that, without loss of
generality, 
in the standard parametrization  $\epsilon$ can be fixed to any value by  a suitable rescaling of the other 
parameters.  Analogously, in the  photonic parametrization, Eq. (\ref{eq:standard_to_photonics}), the
parameters can be rescaled to maintain $\beta J_0$ fixed.


\subsection{Statistical Mechanics of Quenched Disordered Systems}
The couplings in the effective Hamiltonian (\ref{eq:effective_Hamiltonian_3}) are extracted for an appropriate probability distribution
and remain fixed - \emph{quenched} -  in the dynamics.
Then the free-energy density $\phi_N[J]$, as any other observable, 
depends on the particular realization $J$ of the disordered couplings.
For a vast class of observables, including $\phi_N[J]$, however, the
dependence disappears as the system becomes  sufficiently large. \cite{MPVBook}
Such property, called \emph{self-averaging},  implies that 
\begin{align}
 \lim_{N \to \infty} \phi_N [J] = \phi,
\end{align}
where $\phi$ does not depend on $J$
and is equal to the thermodynamic limit $N\to\infty$ 
of the average of  $\phi_N[J]$ over the distribution 
$P[J]$ of $J$:
\begin{align}
 \phi = {\overline{\phi}} =&
 \lim_{N \to \infty} \int \mathcal{D} J \, P[J] \, \phi_N [J]
\\
\nonumber
 =&  - \lim_{N \to \infty} \frac{1}{\beta N}\, \overline{\log Z_N[J]} \, .
\end{align}
where
\begin{eqnarray}
Z_N[J]&=& \int {\cal D} a \; \eu^{- \beta \mathcal{H}[a,J]}
\\
\nonumber
&&{\cal D}a \equiv \prod_{k=1}^N da_k~da^*_k 
\end{eqnarray}
The average of the logarithm of the partition function
 can be performed using the replica trick: \cite{EdwardsAnderson,MPVBook}
one considers $n$ copies of the system and evaluates the replicated partition function
\begin{align}
\label{eq:zrep_ave}
 \overline{Z_N^n} =& \,
 \int D J \; P(J) \, \int {\cal D} a_1 \cdots {\cal D} a_n \;
 \\
 \nonumber
 & \times \eu^{- \beta
  \left[ \mathcal{H}(a_1,J) + \ldots + \mathcal{H}(a_n,J) \right] } 
\end{align}
as function of $n$. A continuation to real $n$ is considered down to values $n<1$, so that the free energy density is eventually obtained as the limit
\begin{align}
 \phi = - \lim_{N \to \infty} \lim_{n \to 0} \frac{1}{\beta N}  \frac{ \overline{Z_N^n}-1}{n} \, .
\end{align}
The replicated partition function $\overline{Z_N^n}$ for large $N$, and fixed $n$,
can be evaluated using the saddle-point method. Therefore, the 
  thermodynamic limit $N \to \infty$ and the limit $n \to 0$ are essentially inverted in the evaluation.
The mathematical foundations of the method are not simple and many efforts have been
necessary to investigate this problem.
In this scenario the well known Replica Symmetry Breaking scheme has been proposed by Parisi \cite{Parisi79, Parisi80}
in the late 70's
and rigorously proved by Guerra \cite{Guerra03} and Talagrand \cite{Talagrand06} about 25 years later.
This Ansatz solves the problem showing a distinctive picture of the underlying structure of the phase space \cite{Parisi84PRL}
and, hence, conferring a key role to the replica trick.

\subsection{Order Parameters in the Replica Formalism}
To apply the replica method to the Hamiltonian  (\ref{eq:effective_Hamiltonian_3})
it is convenient to express first the complex mode amplitude $a_j$ in term of its rescaled real and imaginary part,
respectively $\sigma_j$ and $\tau_j$, as:
\begin{equation}
\label{eq:param}
 a_j = \sqrt{\epsilon} \left( \sigma_j +\iu \tau_j \right) \, , 
 \qquad j= 1, \ldots N .
\end{equation}
The spherical constraint then takes the form
\begin{equation}
\label{eq:spherical_constraint_1}
 \sum_{j=1}^N \left( \sigma_j^2+\tau_j^2 \right) = N  
\end{equation}
while the Hamiltonian (\ref{eq:effective_Hamiltonian_3})  becomes
\begin{align}
\label{eq:effective_Hamiltonian_realImmaginary}
 \mathcal{H} = &- \epsilon\sum_{j<k}  J_{jk} \left( \sigma_{jk}+\tau_{jk} \right) 
 \\
 &
 - \epsilon^2 \sum_{j<k<l<m}  J_{jklm} \left( \sigma_{jklm}+\tau_{jklm} + \varphi_{jklm} \right), 
\nonumber \end{align}
where
we have introduced the short-hand notation 
\begin{eqnarray}
\nonumber
\sigma_{jk} &\equiv& \sigma_j \sigma_k 
\\
\nonumber
\sigma_{jklm} &\equiv& \sigma_j \sigma_k \sigma_l \sigma_m
\end{eqnarray}
 and similarly 
for $\tau$, and

\begin{eqnarray}
 \varphi_{jklm} &=&  \left( \psi_{jk,lm}+\psi_{jl,km}+\psi_{jm,kl} \right) / 3 
 \nonumber
 \\
 \nonumber
 \psi_{jk,lm} &= &\sigma_{jk} \tau_{lm} + \sigma_{lm} \tau_{jk}.
 \end{eqnarray}

In the limit of large $N$ the replicated partition function averaged over the coupling probability distribution
(\ref{eq:disorder})  can be written as an integral  in the replica space of a functional
 of two matrices $\Qp$ and $\Qm$ and two parameters $m_\sigma$ and $m_\tau$,
details can be found in Appendix \ref{averages_quenched_disorder}:
\begin{align}
\label{eq_Z_BZ}
  \overline{Z_N^n} = \int \mathcal{D} \Qp \, \mathcal{D} \Qm \, \mathcal{D} m_\sigma \, \mathcal{D} m_{\tau} \, 
  \eu^{- N G[Q,R,m_\sigma,m_\tau]} \, ,
\end{align}
where 
\begin{align}
 \label{eq:functional_F_general}
 -G[\Qp,\Qm,&m_\sigma,m_\tau] = 
         \frac{1}{2} \sum_{a,b}  g (\Qp_{\a\b},\Qm_{\a\b})  
 +  n\, k(m_{\sigma}, m_{\tau}) 
 \nonumber
 \\
 &+ \frac{1}{2} \ln \det (\Qp+\Qm) 
    - \frac{m_{\sigma}^2}{2} \sum_{a,b} \left( \Qp+\Qm \right)^{-1}_{\a\b} 
 \\
 & + \frac{1}{2} \ln \det (\Qp-\Qm)  
     -  \frac{m_{\tau}^2}{2} \sum_{a,b}    \left( \Qp-\Qm \right)^{-1}_{\a\b}.
     \nonumber
\end{align}
The sums over the replica indexes $\a$ and $\b$ run from $1$ to $n$, and
$g(x,y)$ and $k(x,y)$ are the two-variable functions:
\begin{eqnarray}
\label{eq:definition_g}
  g (x,y) &=&
  \xi_2 (x^2+y^2)   
  + \frac{\xi_4}{2} (x^4 + y^4+ 4 x^2 y^2),  
\\
\label{eq:definition_k}
 k(x, y) &=&
  b_2  \left(  x^2 +  y^2 \right) 
 + b_4  \left(
  x^2 + y^2
  \right)^2.
\end{eqnarray}
The matrices $\Qp$ and $\Qm$ and $m_\sigma$ and $m_\tau$
are related to the  mode amplitudes by
\begin{align}
\label{eq:Q_mat}
 \Qp_{\a\b} &
 = \frac{1}{N}\sum_{j=1}^N  (\sigma^{\a}_j \sigma^{\b}_j + \tau^{\a}_j \tau^{\b}_j)
 =\frac{1}{\cal E} \sum_{j=1}^N  \Re\bigl[a^{\a}_j \left( a^{\b}_j \right)^\ast \bigr] 
  \\
 \Qm_{\a\b} &
 =\frac{1}{N}\sum_{j=1}^N  (\sigma^{\a}_j \sigma^{\b}_j- \tau^{\a}_j \tau^{\b}_j)
 = \frac{1}{\cal E} \sum_{j=1}^N  \Re \left[  a^{\a}_j  a^{\b}_j \right] 
 \label{eq:def_overlap_ReIm}
\end{align}
and\footnote{
The $\sqrt{2}$ is  chosen to have real and imaginary parts of the magnetization ranging 
from $-1$ to $1$ in the case of total energy equipartition among modes, i.e., $|a_j|=1$, for all 
$j$, and between real and imaginary part, i.e.,
$\sigma^2=\tau^2$.
}
\begin{equation}
\label{eq:def_mag}
 m_{\sigma} =  \frac{\sqrt{2}}{N} \sum_{j=1}^N \sigma_j^{\a} , \qquad 
 m_{\tau}      =  \frac{\sqrt{2}}{N} \sum_{j=1}^N \tau_j^{\a},
\end{equation}
\begin{equation}
m_{\sigma}+\iu\, m_{\tau}
         = \frac{\sqrt{2}}{\cal E}\sum_{j=1}^N a^{\a}_j.
\end{equation}
These are the order parameters of the theory.
The ``magnetization'' $m_\sigma$ and $m_\tau$ are single replica quantities
and cannot depend on the particular replica $\a$ because replicas are identical.

In the thermodynamic limit $N\to\infty$ the integral in the replica space can be evaluated using the 
saddle point method, leading to
\begin{equation}
\label{eq:free-en}
\beta \phi = \beta \phi_0 
         + \lim_{n\to 0} \mbox{\rm Extr}\,\frac{1}{n}\,G[\Qm,\Qp,m_\sigma,m_\tau] 
\end{equation}
where $\phi_0$ is an irrelevant constant.
The functional $G[\Qp,\Qm,m_\sigma,m_\tau] $
must be evaluated at its stationary point that, as $n\to 0$, gives the maximum 
with respect to variations of $\Qp$ and $\Qm$ and the minimum with respect to 
variations of $m_\sigma$ and $m_\tau$.

To find the stationary point of $G[\Qp,\Qm,m_\sigma,m_\tau] $ 
an ansatz on the structure of the matrices $Q$ and $R$  is necessary.
It turns out that the simplest ansatz of assuming  $\Qp_{\a\b} = \Qp$ and $\Qm_{\a\b}=\Qm$ 
for all $\a \neq \b$,
 i.e., of assuming that all replicas are equivalent under pair exchange,
is not thermodynamically stable in  the whole phase space, specifically 
at low temperature/high power.
 Therefore, one  must allow for a spontaneous Replica Symmetry Breaking (RSB) and 
 construct the solution accordingly. 
Following the Parisi scheme, \cite{MPVBook}  a 
$n\times n$ matrix $M$ in a $\mathcal{R}$-step RSB state\footnote{
We use the symbol $\mathcal{R}$ instead of the usual $R$ for
number of RSB to avoid possible confusion with the matrix $R_{\a\b}$.}
is
described by $\mathcal{R}+2$ parameters 
$\left( M_{\mathcal{R}+1}, M_\mathcal{R}, \ldots M_0 \right)$
by dividing it along the diagonal into blocks of decreasing linear size $p_r$, with 
$1=p_{\mathcal{R}+1} < p_\mathcal{R}  < \ldots < p_0 = n$, and
assigning $M_{\a\b} = M_r$  if the replicas $\a$ and $\b$ belong to the same block 
of size $p_r$ but to two distinct blocks of size $p_{r+1}$.

Introducing  the $\sigma$ and $\tau$ overlap matrices $A$ and $B$ 
\begin{align}
 \nonumber
 A_{\a\b} =& \Qp_{\a\b} + \Qm_{\a\b}
 = \frac{2}{N} \sum_{j=1}^N  \sigma^{\a}_j \sigma^{\b}_j 
 \, , 
 \\
 \label{eq:definition_matrix_AandB}
 B_{\a\b} =& \Qp_{\a\b} - \Qm_{\a\b} 
  = \frac{2}{N} \sum_{j=1}^N  \tau^{\a}_j \tau^{\b}_j 
 \, ,
\end{align}
the functional $G$ for a generic $\mathcal{R}$-RSB solution is conveniently written as
\begin{align}
\label{eq:freeEnergy_RSB}
 -2 G = &  \sum_{r=0}^{\mathcal{R}+1} (p_r-p_{r+1}) g(\Qp_r,\Qm_r) 
 \\
 \nonumber
 &+ 2 k(m_\sigma, m_\tau) 
 -  \frac{m_\sigma^2}{A_{\widehat{0}}}
 -  \frac{m_\tau^2}{B_{\widehat{0}}} 
 \\
 \nonumber
 & + \log (A_{\mathcal{R}+1}-A_\mathcal{R}) + \sum_{r=1}^{\mathcal{R}+1} \frac{1}{p_r} \log \frac{A_{\widehat{r}}}{A_{\widehat{r+1}}} + \frac{A_0}{A_{\widehat{1}}} 
 \\
 & + \log (B_{\mathcal{R}+1}-B_\mathcal{R}) + \sum_{r=1}^{\mathcal{R}+1} \frac{1}{p_r} \log \frac{B_{\widehat{r}}}{B_{\widehat{r+1}}} + \frac{B_0}{B_{\widehat{1}}},
\nonumber
\end{align}
where the hatted quantities are the Replica Fourier Transform (RFT) 
of a $\mathcal{R}$-RSB matrix $M_r$  
defined as\cite{ParisiFourier,ReplicaFourier97, CrisantiRFT}
\begin{align}
& M_{\widehat{k}} =  \sum_{r=k}^{\mathcal{R}+1} p_r \left( M_r-M_{r-1} \right)  
\nonumber
\\
 &
 M_r = \sum_{k=0}^r \frac{1}{p_k} \left( M_{\widehat{k}}-M_{\widehat{k+1}} \right),
\label{eq:def_RFT}
\end{align}
where it is intended that terms with indices outside the respective interval of definition are null.

The solutions of the stationarity equations, see  
Appendix \ref{App:Solutions}, are either of the form 
\begin{equation}
 \label{eq:m_choice}
 m_\tau = 0 \ \text{and}\  B_r = 0 \to \Qm_r = \Qp_r,  \quad
 r=0, \ldots \mathcal{R},
\end{equation}
or  (because of the symmetry $\sigma \leftrightarrow \tau$):
\begin{equation}
\label{eq:mt-sol}
 m_\sigma = 0 \ \text{and}\  A_r = 0 \to \Qm_r = -\Qp_r,  \quad
 r=0, \ldots \mathcal{R}.
\end{equation}
In the following, without loss of generality, we consider the first form, Eq. (\ref{eq:m_choice}), and,
for simplicity, we drop the subscript in the magnetization writing
 $m_\sigma = m$.

\section{Optical Regimes and Phase Diagram}
\label{sec:results}
The phase diagram obtained from the analysis of the solution of the mean-field model
consists of four different phases distinguished by the values
of the order parameters $Q$, $R$ and $m$:

\begin{itemize}
 \item \emph{Incoherent Wave (IW):}
        replica symmetric solution with all order parameters equal to zero. 
        The modes oscillate incoherently in this regime and the light is emitted in the form of a continuous wave. 
        At low pumping, else said high temperature,
       this is the only solution. 
       It corresponds to the Paramagnetic phase in spin models.
 \item \emph{Standard Mode-Locking Laser (SML):}
       solutions with $m\not=0$, with or without replica symmetry breaking.
       The modes oscillate coherently with the same phase and the light is emitted in form of optical pulses.
       It is the only regime at high pumping (low temperature) when $b_2$ and $b_4$ are large with respect to $\xi_2$ and $\xi_4$, that is when the degree of disorder $R_J$ is small.
       It corresponds to the Ferromagnetic phase in spin models.
 \item \emph{Random Laser (RL):}
       the modes do not oscillate coherently in intensity, so that $m=0$, 
       but $R_{aa} = \langle \sigma^2 \rangle - \langle \tau^2 \rangle \neq 0$ implying a phase coherence
       and the overlap matrices have a nontrivial structure.
       It is the only phase in the high pumping limit for large disorder, i.e.,  when $\xi_2$ and $\xi_4$ are large  with respect to $b_2$ and $b_4$.
       It corresponds to the Spin Glass phase in spin models with disordered interactions. 
 \item \emph{Phase Locking Wave (PLW):}
       all order parameters vanish but $R_{aa}$, so that $\langle \sigma^2 \rangle \neq \langle \tau^2 \rangle $, signaling a locking of the mode phases in a specific direction.
       This regime occurs in a region of the phase space intermediate between the IW and RL regimes
      if $\xi_4 > 0$ ($R_J>0$).
       This new thermodynamic phase, to our knowledge never observed in spin models,
        follows from the peculiar kind of spins considered, displaying both a phase and a magnitude,
       that leads to a combination of  XY (only phase) and real spherical 
       (only magnitude) spins.
       The locking in the two degrees of freedom of each 
       complex amplitude does not happen concurrently as the
        temperature is lowered in presence of (even a small amount of) 
        disorder. Mode phases lock first, in what we call the PLW regime. 
\end{itemize}

The narrow band approximation, in which the present model is exact,
makes the study of the pulsed emission particularly delicate.  Indeed,
if the magnetization is nonzero, as in the SML phase, the light is
generated in form of short pulses when the presence of different
frequencies in the spectrum is considered and this property cannot be reproduced in the narrow-band limit.  Besides, we note that a
ML phase with a nonzero phase delay, as described in Ref.
[\onlinecite{Antenucci14b, Marruzzo14}], associated with pulsed emission in an
unmagnetized state is not feasible, as the frequency does not play any
role in the mode evolution. Nevertheless, it is possible to analytically solve the model in the whole phase diagram and describe all regimes for optically active media.

Different replica symmetry breaking solutions, and corresponding RL regimes,  
occur by varying the ratio $r \equiv \xi_2 / \xi_4$, or, equivalently,
varying the strength of disordered nonlinearity $\alpha$.  In particular, for
$r$ large enough ($\alpha$ small enough) the stable thermodynamic phase is
expected to be characterized by a Full Replica Symmetry Breaking (FRSB) 
state.\cite{CrisantiLeuzzi_sp_NPB}  In realistic optical systems the
$2$-body interaction is usually not dominant above the lasing threshold, 
cf. discussion in Sect. \ref{sec:derivation_of_the_model}.
However,  there can be systems where the
damping due to the openness of the cavity is strong enough to compete
with the non-linearity and a FRSB state may emerge. In this case the transition turns out to be continuous in the
order parameters: the overlap is zero at the critical point and grows continuously as the power is increased above threshold.
In this work we shall limit ourselves to the analysis of the first step of 
replica symmetry breaking, by considering solutions with only one step of replica symmetry 
breaking (1RSB). We, hence, provide the exact mean-field laser solution for $r$ low
enough (specifically, $r < 0.517 $ for $ b_2 = b_4 = 0 $). Due the continuous nature of the transition to the FRSB regime, though, the 1RSB solution is also a
reliable approximation of the lasing solution not far from threshold
for larger $r$.

\subsection{RS and 1RSB phases}

Solutions with non-null magnetization $m$ originates because of the ordered, non-random,  
part of the Hamiltonian.
These are more easily described by
introducing a suitable external field $h$
and  considering the fully-disordered model, i.e., with $J_0^{(p)}=0$,  
with Hamiltonian
\begin{equation}
\label{eq:ham_with_ext_field}
 \mathcal{H}_h  = \mathcal{H} - \sqrt{2} h \sum_j \sigma_j,
\end{equation}
where $\mathcal{H}$ is given in  (\ref{eq:effective_Hamiltonian_realImmaginary})
and, the $\sqrt{2}$ follows from the definition \eqref{eq:def_mag} of 
$m_\sigma$. 
Absorbing the factor $\beta$ into the field writing $b = \beta h$, 
the solutions of the two models are identical provided 
\begin{equation}
 b = \frac{\partial }{\partial m} k(m,0),
\end{equation}
so that we have the \emph{unfolding equation}
\begin{equation}
\label{eq:unfolding_equation}
b =  \left( 2 b_2 + 4 b_4 m^2 \right) m,
\end{equation}
which relates the $m\not=0$ states of the original model (\ref{eq:effective_Hamiltonian_3}) 
to the ones of  model (\ref{eq:ham_with_ext_field}). 
For the solution with  $m_\sigma=0$ and $m_\tau=m$  there is an equivalent mapping.
The origin of this connection is discussed in Appendix \ref{App:Solutions}.
Note that for $b\not=0$ all solutions have $m\not=0$, and hence correspond to the SML regime.
The other phases with $m=0$ correspond to the $b=0$ case. 
The appearance of nontrivial solutions of the unfolding equation (\ref{eq:unfolding_equation}) 
then  signals  the transition to the SML regime.

The parameterization $a\to (\sigma,\tau)$ given in \eqref{eq:param} 
ensures that the diagonal elements $\Qp_{\a\a}$ are equal to $1$.
The diagonal elements of $\Qm_{\a\a}$  are also independent from the replica index 
$a$ but the value depends on the control parameters.
The  analysis of the solution \eqref{eq:m_choice} is then simplified by introducing the ``scaling factor"
\begin{equation}
\overline{a} = \Qp_{\a\a} + \Qm_{\a\a} = 1 + \Qm_{\mathcal{R}+1}, 
\end{equation}
varying between $1$ and $2$, 
and using the rescaled overlap parameters $q_r$ defined as
\begin{equation}
  \Qp_r + \Qm_r = 2 \Qp_r = \overline{a}\, q_r,
  \quad r = 0,\dotsc,\mathcal{R}.
\end{equation}
For the solution \eqref{eq:mt-sol} one has a similar parameterization.

In this work we shall only consider solutions with $\mathcal{R}=0$ (RS) and 
$\mathcal{R}=1$ (1RSB). The study of solution with more complex RSB structure 
will not be reported here.

The RS solution is described by the three parameters
$q_0$, $\overline{a}$  and $m$ and,  neglecting all unnecessary terms, the free energy 
functional \eqref{eq:free-en} reads:
\begin{align}
\label{eq:free_energy_RS}
 2 \beta \phi (q_0,&\overline{a},m) = 
 - \overline{g}
 + g_0
 - \frac{q_0 - m^2 / \overline{a}}{1-q_0}
 \nonumber \\
 &
 - \log \left[ \overline{a} \left( 2-\overline{a} \right) (1-q_0) \right]
 - 2 b m,
\end{align}
where we have used the short-hand notation:
\begin{equation}
 \begin{split}
  \overline{g} = & g \left( 1,\overline{a}-1 \right), \\
   g_{r} = & g \left( \frac{\overline{a}}{2} q_{r} , \frac{\overline{a}}{2} q_{r} \right).
  \end{split}
\end{equation}
Stationarity of $\phi(q_0,\overline{a},m)$ 
with respect to variation of $q_0$, $\overline{a}$ and $m$ leads to
the stationary equations:
\begin{align}
  \label{eq:RS_stationarity_field}
  \nonumber
  \overline{a}\, \Lambda_0 =&  
 \frac{q_0}{(1-q_0)^2} - \overline{a} \, b^2  \, ,
 \\
  \overline{a}\, \overline{\Lambda} - \overline{a} \Lambda_0 
  =& - \frac{1}{1 - q_0} +  \frac{\overline{a}}{2-\overline{a}}  \, ,
\end{align}
and  $m = b\, \overline{a}(1-q_0)$,
where
\begin{equation}
 \begin{split}
 \overline{\Lambda} = & \Lambda \left( \overline{a}-1,1 \right),
\\
 \Lambda_{r} = & \Lambda \left( \frac{\overline{a}}{2} q_{r} , \frac{\overline{a}}{2} q_{r} \right),
  \end{split}
\end{equation}
with
\begin{equation}
\label{eq:Lambda}
\Lambda \left(x, y \right) = \frac{\partial}{\partial x} g(x,y) = 
2  x \left[ \xi_2 + \xi_4 \left( x^2+ 2  y^2 \right) \right].
\end{equation}

The 1RSB solution
requires five parameters: $q_0$, $q_1$,  $\overline{a}$, $m$ and
the ``block size'' $p_1$. The latter  becomes a real number  for $n\to 0$, the RSB parameter
$p_1=x\in[0,1]$.
The free energy  for the 1RSB  solution then reads:
\begin{align}
\label{eq:free_energy_1RSB}
\nonumber
 2 \beta \phi(q_0,q_1,&x,\overline{a},m) =  
 - \overline{g} + (1-x) g_1  + x  g_0   
 \\
 &
 -  \frac{  q_0 -m^2/\overline{a} }{q_{\hat{1}}} 
 - \log \left[  \overline{a} (2-\overline{a}) (1- q_1)  \right] 
 \nonumber 
 \\
 &
  - \frac{1}{x} \log \frac{q_{\hat{1}}}{q_{\hat{2}}}
  - 2 b m,
\end{align}
where
\begin{align}
 q_{\hat{2}} &= 1 -q_1 \, , &  q_{\hat{1}} = 1-q_1 + x (q_1-q_0) \, .
 \label{eq:definiton_parameters_chi}
\end{align}
Stationarity of the functional with respect to variations of 
$q_0$, $q_1$,  $\overline{a}$, $m$ gives the stationarity equations:
\begin{align}
\nonumber
 \overline{a} \Lambda_0  =&  
 \frac{q_0}{q_{\hat{1}}^{\, 2}} - \overline{a} \, b^2  \, ,
 \\
 \overline{a} \Lambda_ 1 
 - \overline{a} \Lambda_0 
  =&   \frac{q_1 - q_0}{q_{\hat{2}}\, q_{\hat{1}} },
  \label{eq:stat_1RSB}
 \\
 \overline{a}\, \overline{\Lambda}  - \overline{a} \Lambda_1 
  =& - \frac{1}{1-q_1} +  \frac{\overline{a}}{2-\overline{a}} ,
  \nonumber
  \\
m =& \overline{a} \, b \, q_{\hat{1}}.
\nonumber
\end{align}
The role of the parameter $x$ is more subtle. If the free energy functional is required to be
stationary also with respect to variations of $x$, one then obtains the additional stationarity equation:
\begin{align}
   g_1    - g_0 
   =& - \frac{1}{x^2} \log \frac{q_{\hat{2}}}{q_{\hat{1}}} 
\label{eq:stat_x_1RSB}
   \\
\nonumber    
& - (q_1-q_0) \left[ \frac{1}{x q_{\hat{1}}} -  \frac{q_0-m^2 / \overline{a}}{q_{\hat{1}}^{\, 2}} \right],
\end{align}
which describes the \emph{static} solution with null complexity, see
Sect. \ref{sec:complexity} for more detail. 
This equation can be rewritten, making use of the other Eqs. (\ref{eq:stat_1RSB}), as
\begin{equation}
\label{eq_1RSB_h}
     2 \frac{g_1 - g_0 - \overline{a} \Lambda_0 (q_1-q_0)}
     {\overline{a} (\Lambda_1 - \Lambda_0) (q_1-q_0)} = z(y),
\end{equation}
where $y = q_{\hat{2}}/q_{\hat{1}} \in [0,1]$ and
\begin{equation}
\label{eq:def_zFunction}
 z(y) = -2 y \frac{1-y+ \ln y}{(1-y)^2},
\end{equation}
is the CS $z$-function.\cite{Crisanti92} 
This form is particularly useful for the numerical solution of the stationary equations, in particular to 
find solutions for fixed value of $x$ (that is, along a $x$-line).
Note also that it  depends only on the ratio $r = \xi_2/\xi_4$
and not on $\xi_2$ and $\xi_4$ separately.

Alternatively, the value of $x$ can be fixed by requiring that the complexity, i.e., the number of 
equivalent metastable states - or ergodic components - is maximal. This choice leads to the
{\sl dynamic} solution, and to the following equation for $x$:
\begin{equation}
\label{eq:x_dyn}
 \left.
                 \left[ \frac{\partial}{\partial x} + \frac{\partial}{\partial y} \right]\, \Lambda(x,y)
                 \right|_{x = y= \overline{a} q_1 / 2} 
                  = \frac{2}{\overline{a}^{\, 2}\,(1-q_1)^2}.
\end{equation}
We defer a more detailed discussion to  Sect. \ref{sec:complexity}.

The analysis of the solutions can be done in two steps. First we discuss the solutions 
in presence of 
a fixed field $b = \beta h$. Then, once the different phases have been identified, we shall 
{\sl unfold} $b$ in terms of $m$ using the unfolding equation to recover  the original
RL problem. 
For this reason the phase diagrams will be reported in the 
$(\xi_2,\xi_4,b)$, $(\xi_2,\xi_4, b_2, b_4)$ and
$({\cal P}, \alpha, \alpha_0, R_J)$ spaces, as appropriate.
The relation among the different spaces are given in Eqs.
\eqref{eq:standard_to_photonics} and the unfolding equation (\ref{eq:unfolding_equation}).

\begin{figure}[t!]
\begin{center}
\includegraphics[width= 0.99 \columnwidth ]{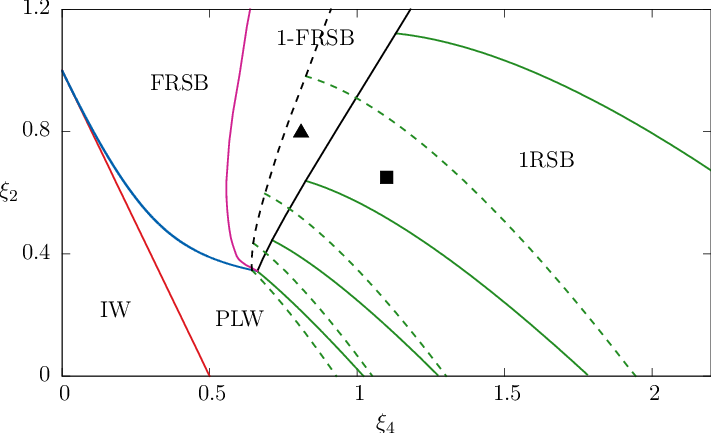}
\caption{ Phase diagram $\xi_2$, $\xi_4$ for $b=0$ with static (solid
  green) and dynamic (dashed green) $x$-lines.  In the 1RSB phase,
  $x$-lines with different value of $x$ are shown both in the static and dynamic cases:
  $x= 0.45, 0.6, 0.8$ and $x=1$  top to bottom.  The $x=1$
  static $x$-line is the static glassy RFOT  line between the PLW phase and the RL with 1RSB
  phase.  The $x=1$ dynamic $x$-line is where the dynamics arrests 
  because of the exponential number of metastable states characteristic of the RFOT.
 The solid black  line marks the end point of the $x$-lines and a 1-FRSB appears.
  The dashed black line is the analogous critical line of dynamic $x$-lines.
  Both static and dynamic 1-FRSB phases end on the
  solid magenta line. Here the complexity vanishes and 
 a  transition  to a FRSB  phase occurs. 
 Finally the solid blue line marks the direct transition between the PLW and the FRSB phases.  
 The order parameters are continuous crossing this line. 
   The $\blacksquare$ and $\blacktriangle$ symbols correspond to the positions of the data shown in Fig. \ref{fig:complexity_eigenvalues_newStable}.
  }
\label{fig:dynamic_lines}
\end{center}
\end{figure}


\begin{figure}[t!]
\begin{center}
\includegraphics[width= 0.99 \columnwidth ]{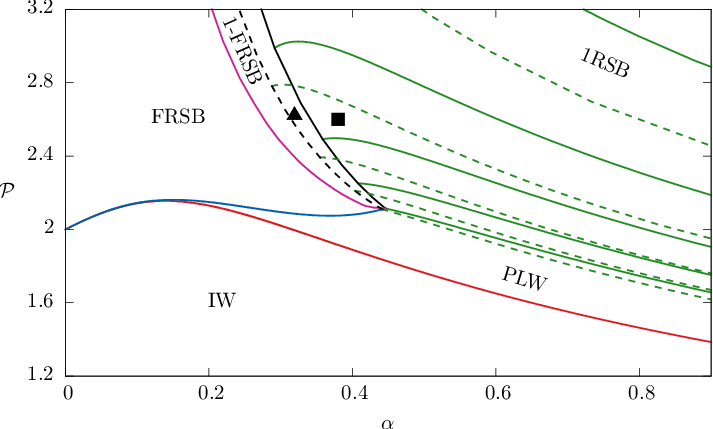}
\caption{ Phase diagrams for $b=0$ in the photonic parameters, pumping
  rate ${\cal P}$ and degree of openness $\alpha$, for $R_J = \beta
  J_0 = 1$.  }
\label{fig:dynamic_lines_PP}
\end{center}
\end{figure}

\subsection{Phase Diagram at Fixed Field}

To study the transition to the RL state we consider first the case of zero external field $h$.
The complete phase diagram for $b=0$, including phases with RSB structure 
more complex than RS or 1RSB,  is shown in Figs. \ref{fig:dynamic_lines}
and \ref{fig:dynamic_lines_PP}.
In particular, the IW is the only phase at high enough temperature. Lowering the temperature, for $
\xi_4>0$, the IW becomes metastable as the PLW phase appears continuously. 
The case of  $\xi_4=0$ is different: increasing $\xi_2$ the IW phase becomes unstable and a
 transition occurs to  replica symmetric  RL  phase.

For $r = \xi_2 / \xi_4 < 0.517$, i.e., $\alpha>\alpha_{\rm nl}\simeq (3-1.76382 \epsilon) (3-1.03703 \epsilon^2)^{-1}  $, 
increasing $\xi_4$ 
the PLW regime undergoes a Random First Order Transition (RFOT) \cite{RFOT} to a glassy RL phase with 1RSB:
a jump is present in the order parameters $\Qp$ and $\Qm$ but the internal energy remains 
continuous.  Thus,  no latent heat is exchanged.

For $r = \xi_2 / \xi_4 > 0.517$ the PLW becomes unstable at a critical temperature 
where the transition occurs to a RL regime with FRSB, of the kind reported in Ref. [\onlinecite{CrisantiLeuzzi_sp_NPB}].

\begin{figure}
\begin{center}
\includegraphics[width= 0.99 \columnwidth ]{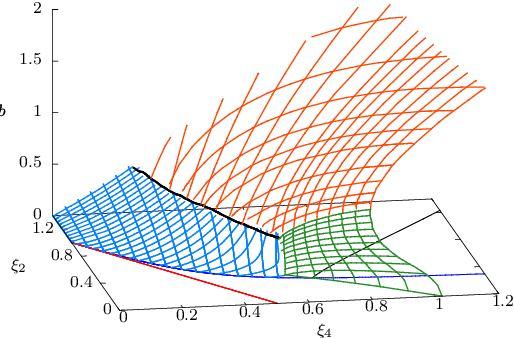}
\caption{
  Phase diagram of the  $2+4$ complex spherical model at fixed external
  field in the $(\xi_2,\xi_4,b)$ space.  Only the static solution is shown.
  The green surface, given by  the 1RSB solution with $x=1$,  
  is the critical RFOT surface between 
  the RS and 1RSB phases.  
  The red surface, given by the 1RSB solution with  $t=1 \to q_1 = q_0$,
   gives critical surface of the continuous transition between the
  RS and 1RSB phases.  
  The black line marks the end-line of the the 1RSB solution.  
  The blue surface is critical surface where the (continuous) 
  transition between the RS to FRSB phases
  occurs.}
\label{fig:PhDi_P_alpha_h}
\end{center}
\end{figure}

The complete phase diagram at non-zero external field is shown in Fig. \ref{fig:PhDi_P_alpha_h}.
In this case  it is useful to introduce the ratio $t =q_0/q_1\in[0,1]$.
%
%
The 1RSB solutions can be found fixing the parameters $r, \, x, \, t$ and using 
the stationary point equations to find the value of $\xi_4$, $b$, $y$ and $\overline{a}$.
Two surfaces are of particular interest: the surface $x=1$, corresponding to the RFOT between the RS and 1RSB solution (Fig. \ref{fig:PhDi_P_alpha_h}: green surface), 
and the surface $t=1 \leftrightarrow q_0 = q_1$ corresponding to the continuous transition between the RS and RSB phase (Fig. \ref{fig:PhDi_P_alpha_h}: red surface).

A condition for the existence of the 1RSB solution can be derived considering the stationarity equations for $t$ close to $1$.
In this case the ratio $w \equiv (1-t)/(1-y)$ has the finite limit
\begin{align*}
 w = -\frac{-9 \overline{a}^2 +8 r x + 3 \sqrt{9 \overline{a}^4-16 \overline{a}^2 r x^2-16 \overline{a}^2 r x}}{8 r x^2} \, .
\end{align*}
as $t\to 1$ and $y\to 1$ simultaneously.
The 1RSB solution thus exist only if
\begin{align*}
 16 rx (x+1) \leq 9\, \overline{a}^2 \, .
\end{align*}
This critical line is drawn in black in Fig. \ref{fig:PhDi_P_alpha_h}.
Since  the condition becomes more and more stringent increasing $x$, 
this line does not exist for $r< 9/32$.
The line lies on the RS instability surface,
drawn in blue in Fig. \ref{fig:PhDi_P_alpha_h}, 
where the eigenvalue
\begin{align}
  \leftidx{_{2}}\Lambda(0;1,1) 
= -2 \xi_2 - \frac{9}{2} \xi_4 \overline{a}^2 q_0^2 + \frac{2}{\overline{a}^2 (1-q_0)^2},
 \label{eq:eigenvalues_RS_text_field}
\end{align}
of the fluctuations about the RS saddle point, with 
$\overline{a}$ and $q_0$ are evaluated at RS saddle point \eqref{eq:RS_stationarity_field},
vanishes.
The eigenvalue can be computed by extending the calculation of Appendix \ref{app:RS_Fluct}
to the case $Q_0\not=0$. Alternatively it can be derived from the 1RSB stability 
analysis of Appendix \ref{app:1RSB_Fluct} by setting $x=0$ and replacing $Q_1$ 
of the 1RSB solution with $Q_0$ of the RS solution. The eigenvalue
\eqref{eq:eigenvalues_RS_text_field} then follows from  \eqref{eq:E7}.

As for the zero-field case, the RS solution is stable in the whole
phase space.  For high
$r$ the 1RSB solution disappears for 
values of $x$ smaller than a threshold value that decreases as $r$
increases and it is replaced by  a FRSB solution,
as it occurs for real spherical spin
models. \cite{CrisantiLeuzzi_sp_NPB}

\begin{figure}
\begin{center}
\includegraphics[width= 0.99 \columnwidth ]{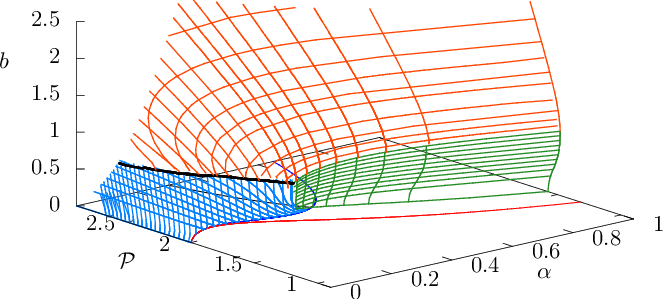}
\caption{ 
           Same as in Fig. \ref{fig:PhDi_P_alpha_h} in the photonic
           $({\cal P}, \alpha, b)$ diagram.
  }
\label{fig:PhDi_P_alpha_h_PP}
\end{center}
\end{figure}

\subsection{Complexity}
\label{sec:complexity}

Looking at Fig. \ref{fig:dynamic_lines} one sees that 
when $b=0$ and $r = \xi_2 / \xi_4 < 0.517 $
the static transition line, the solid $1$-line, from the  PLW to the RL regime is
anticipated by a dynamic transition line, dashed $1$-line. 
Approaching this line from the PLW side a critical slowing down, followed by 
a dynamic arrest on the line, is observed in the time correlation functions.
This behavior, typical of glassy systems, is due to the breaking of ergodicity of the phase space.
The phase space breaks down into a large number $\mathcal{N}$,
increasing {\sl exponentially} with the size of the system, of degenerate metastable
states that dominate the dynamical behavior 
before the true thermodynamic (static) transition is reached, see, e.g.,
Refs. [\onlinecite{Kirkpatrick87}], [\onlinecite{Crisanti93}] and [\onlinecite{Goetze09}].

Lasing in random media is hence expected to
display a {\it glassy} coherent behavior.  With glassy we mean that (i)
a sub-set of modes out of an extensive ensemble of localized passive
modes are activated in a non-deterministic way
and (ii) the whole
set of activated modes behaves cooperatively and belongs to one state
out of (exponentially) many possible ones.
We stress that in characterizing the glassy behavior, by \emph{non-deterministic} we do not mean
deterministic chaos, i.e., high sensitivity to
initial conditions.  Chaos is, actually, a dynamic phenomenon
occurring in laser systems, \cite{Weiss91} but it is independent from
possible glassy properties.  In other words, chaos can occur in these systems, but it does not affect the presence or
absence of glassiness, as, e.g., shown in Ref.
[\onlinecite{Crisanti96}]: it is not a necessary nor a sufficient
feature for random lasers.

The role of the  metastable states can be 
captured by introducing the  \emph{complexity}, aka \emph{configurational entropy},
\begin{align}
 \Sigma \equiv \frac{1}{N}\log \mathcal{N}.
\end{align}
The complexity $\Sigma$ as function of the free energy of the metastable states can
obtained from the  Legendre transform of  $\phi(x)$:\cite{Monasson95,Mezard99,Mueller07}
\begin{eqnarray}
\label{eq_x_complexity}
     &\Sigma (f) + \beta x \phi (x) = \beta x f,
\\
\label{eq:x_f}
 &f = \dfrac{\partial x \phi(x)}{\partial x}
\end{eqnarray}
where $f$ is the free energy of a single metastable state and
$x=x(f)$ is the solution of Eq. \eqref{eq:x_f}.

When $h=0$ the overlap $q_0$ vanishes and the 1RSB free energy functional 
\eqref{eq:free_energy_1RSB}  evaluated on the stationary point
\eqref{eq:stat_1RSB}
reduces to:
\begin{align}
\label{eq:free_energy_1RSB_0}
\nonumber
 2 \beta \phi(x) =   - \overline{g} &+ (1-x) g_1 
 - \log \left[  \overline{a} (2-\overline{a}) (1- q_1)  \right] 
 \nonumber 
 \\
 &
  - \frac{1}{x} \log \frac{q_{\hat{1}}}{q_{\hat{2}}},
\end{align}
which replaced  into the Legendre transform \eqref{eq_x_complexity}, yields
\begin{equation}
 \label{complexity_noField}
\Sigma =
 \frac{\beta}{2}\left[ -\frac{q_1 x}{q_{\hat{1}}}- g_1 x^2
     +\ln \frac{q_{\hat{1}}}{q_{\hat{2}}} \right],
\end{equation}
where $q_1=q_1(x)$.
To obtain the  complexity $\Sigma(f)$ one should replace 
$x=x(f)$ or, alternatively, use $x$ or $q_1$ as a free parameter  and
evaluate $f$ from Eq. \eqref{eq:x_f}.

Stationarity of $\Sigma(f) - \beta xf$ with respect to variation of $f$ 
for fixed $x$ leads to $x = \beta^{-1}(\partial /\partial f)\Sigma(f)$.
Physically acceptable solutions $f$ must have $x\in[0:1]$ and, hence, 
$\Sigma(f)$  is a non-decreasing function of $f$.

The static solution corresponds to the metastable states with the lowest free energy
and null complexity, i. e.,  the lowest acceptable value of $\Sigma$.
For these states $f = \phi$ and from Eq.  \eqref{eq:x_f} one easily recovers
the static stationary condition $\partial_x \phi(x) = 0 $.
The dynamical behavior of the system is dominated by the large number of 
metastable states with the highest physically acceptable free energy $f$, 
for which the complexity reaches its maximum value. 

The range of allowable $f$ is fixed by stability condition of the saddle point replica 
calculation, i.e., that
the two relevant eigenvalues, see Appendix \ref{app:1RSB_Fluct}:
\begin{align}
 \leftidx{_1}\Lambda (0;1,1)
 =& - 2\xi_2 + \frac{2}{\overline{a}^2 [1 -  q_1 (1-x) ]^2},
 \\
  \leftidx{_1}\Lambda (1;2,2)  
 = & 
    -2 \xi_2  - \frac{9}{2} \, \xi_4  \, \overline{a}^2 q^2_1
   + \frac{2}{\overline{a}^{\, 2}(1 -  q_1)^2},
\end{align}
of the fluctuations about the 1RSB saddle point are non-negative.
The eigenvalue  $\leftidx{_1}\Lambda(0;1,1)$ controls the fluctuations with
respect to $q_0=0$ and its vanishing marks the end of the 1RSB phase and the 
appearance of a 1-FRSB phase.\cite{Crisanti04,Crisanti06}
The eigenvalue $\leftidx{_1}\Lambda(1;2,2)$ controls the 
stability with respect to fluctuations of $q_1$. It can be shown that 
it also controls the critical slowing down of the dynamics, \cite{Crisanti07} and, hence, its vanishing 
leads to the {\sl marginal condition} for the arrested dynamics.
The requirement of a maximal complexity is then equivalent to 
$\leftidx{_1}\Lambda(1;2,2)=0$, cf. Eq. \eqref{eq:x_dyn}.

\begin{figure}[t!]
\begin{center}
\includegraphics[width= 0.99 \columnwidth ]{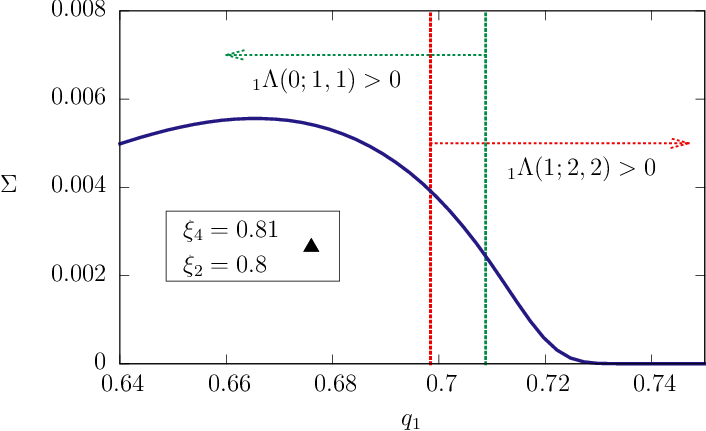}
\includegraphics[width= 0.99 \columnwidth ]{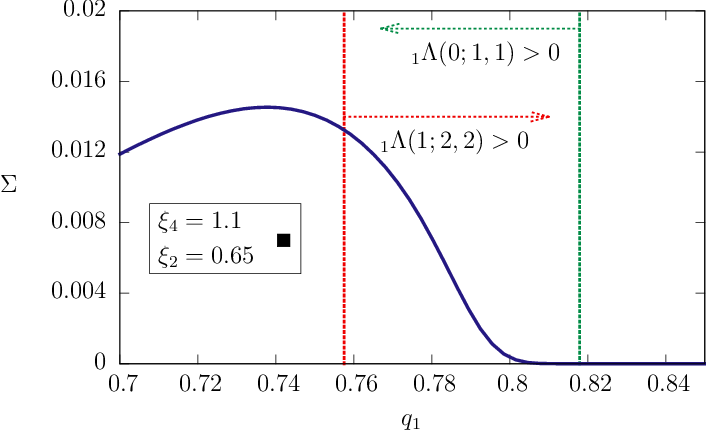}
\caption{
Complexity $\Sigma$ (blue solid line) as function of $q_1$ for 
$\xi_4=0.81$, $\xi_2=0.8$ (upper panel, $\blacktriangle$ symbol in Fig. \ref{fig:dynamic_lines}) 
and  
$\xi_4=1.1$, $\xi_2=0.65$ (lower panel, $\blacksquare$ symbol in Fig. \ref{fig:dynamic_lines}).
The red dashed line indicates the region where the eigenvalue 
$\protect\leftidx_{1}\Lambda(1;2,2)$ is positive. 
The green dashed line indicates the region where the eigenvalue 
$\protect\leftidx_{1}\Lambda(0;1,1)$ is positive. 
The interval of $q_1$ where they are both positive identify the region where 
the 1RSB solution is stable for the given $\xi_2,\xi_4$.
In the upper panel the static 1RSB solution is unstable, $\Sigma=0$ lies outside the 
allowed region, and only the dynamic 1RSB solution exists. The static solution here exists 
but with a more complex RSB structure.
}
\label{fig:complexity_eigenvalues_newStable}
\end{center}
\end{figure}
In Fig. \ref{fig:complexity_eigenvalues_newStable} 
 the form of $\Sigma(q_1)$ is shown for two representative cases.
In the upper panel only the dynamic 1RSB solution exists. The minimum $\Sigma=0$ lies indeed
outside the stability region of the 1RSB  phase and the static  1RSB solution is unstable.
Here the static solution is replaced by a solution with a more complex
RSB structure, namely a 1-FRSB phase.


Decreasing $\xi_2$ reduces the stability region of the 1RSB solution and eventually also the 
dynamic 1RSB solution becomes unstable. This occurs on the dashed black line shown in  
in Fig. \ref{fig:dynamic_lines}. Beyond this line only a 1-FSRB phase, both static and dynamic,
exists.  We shall not enter into the detail of this phase, it is enough to say that,
as it occurs for the 1RSB phase, they differ in complexity.  A further decrease of $\xi_2$ 
reduces the complexity of the dynamic solution and it eventually vanishes when the
magenta solid line is reached. Here the distinction between static and dynamic solution 
disappears and both solutions merge into a FRSB state.

\subsection{Unfolding the field}

In the low temperature phase all solutions with null external field $b$ described so far
belong to the SML phase  $m\not=0$, 
since  $m = b\, \overline{a}q_{\hat{1}}$.
The description of the SML phase in terms of $b$ is then trivial.
To switch to the description in terms of the parameters $b_2$ and $b_4$, 
cf. Eq. \eqref{eq:definition_parameters_b_xi_mu}, we have to unfold the 
value of the field $b$ in terms of  $b_2$ and $b_4$ using the 
\emph{unfolding  equation} \eqref{eq:unfolding_equation}.
The unfolding equation admits the trivial solution $m=0$ and a non-trivial $m\not=0$ solution
if 
\begin{equation}
\label{eq:SML_cond}
2 b_2 + 4 b_4
  m^2 = \frac{1}{\overline{a}\, q_{\hat{1}}}.
\end{equation}
The SML phase exists only in the regions of the phase space where this condition is satisfied. 
As in ordinary ferromagnetic first order transition, the SML  phase may appear 
discontinuously with a finite value of $m$ and
higher  free energy on the spinodal line. 
If this happens the SML phase becomes thermodynamically favorable only
when the SML free energy  becomes equal to the free energy of the $m=0$ solution.
This condition defines the transition line in the phase space.
Since the two phases have equal free energy at the transition they coexist
 and latent heat is exchanged during the transition.

Using the stationary point equations \eqref{eq:stat_1RSB}
and defining $\gamma = b_2 /b_4$, from Eq.  \eqref{eq:SML_cond} one easily gets:
\begin{equation}
\label{eq:b4_q0}
 b_4 =  \frac{1}{\overline{a}\,q_{\hat{1}} } \, 
     \frac{1}{2 \, \gamma + 4 \, b^2 \, \overline{a}^{\,2} q_{\hat{1}} },
\end{equation}
which relates the value of $b_4$ for fixed $\gamma$ to the that of the parameters of the
model with fixed $b$. 
With this parametrization the SML spinodal line corresponds to the minimum values of $b_4$ 
where the $m\neq 0$ solution first appears. 
Since when $m=0$ the field $b$ vanishes and consequently $q_0=0$, it is useful to use $q_0$ as 
free parameter in the model with fixed $b$.  The SML spinodal line for fixed $\gamma$ then 
occurs at  $ b_4^\star =  \min_{q_0} b_4(q_0)$.
The value of $b_4^\star$ is strictly positive, and  a rough estimate gives the bound:
\begin{align*}
 b^\star_4 \geq 
 \left[ 2(1+\mathcal{R})(\gamma +2 \, \epsilon) \right]^{-1}.
\end{align*}
Note that the bound decreases with the number of steps $\mathcal{R}$ of RSB, thus SML phases with higher replica symmetry
breaking steps appear first, if they exists. In the following we shall only discuss the
RS SML phase with $\mathcal{R}=0$. While the discussion of SML phases with more complex
RSB structures, though feasible,  will not be presented here.

For large values of $\gamma$ 
the minimum of $b_4(q_0)$ occurs at $q_0=0$ and the SML appears  continuously with $m=0$.
The spinodal line then coincides with the transition line and 
the transition between the IW/PWL and SML phases is continuous in $m$, with
no phase coexistence and latent heat. 

For sufficiently small $\gamma$ the minimum of $b_4(q_0)$ moves to $q_0>0$ and 
the SML phase appears discontinuously with $m\not=0$. 
Thermodynamic transitions between the IW or PWL  phases and the SML phase occur
at the critical value $b_4^c > b_4^\star$  where the free energy of the phases becomes equal.

\begin{figure}
\begin{center}
\includegraphics[width=0.9\columnwidth]{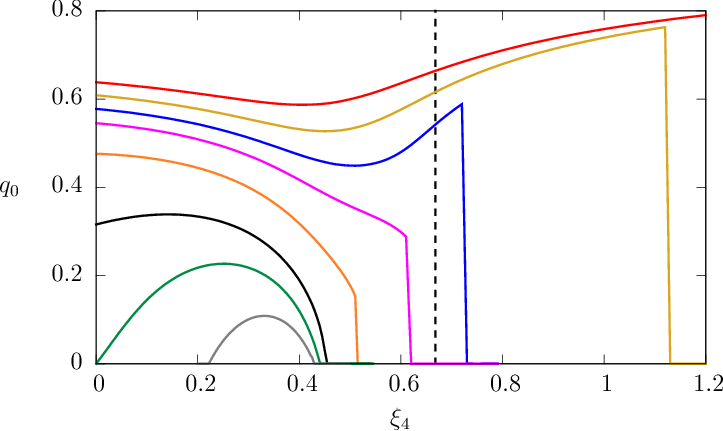}
\caption{ Curves of $q_0$ where $b_4(q_0)$ is minimal 
  as function of $\xi_4$ for fixed $\gamma$ and $r=0.5$
   Bottom to top:  $\gamma = 10^3 ,
  \, 4, \, 2.5 , \, 1.5, \, 1 , \, 0.75, \, 0.5,\, 0.25$.  
  The vertical dashed line marks the value $\xi_4 \simeq  0.67$
  where the 1RSB solution for  $r=0.5$ appears.
 For $\gamma$ small enough the spinodal line of the RS SML phase enters into
  the 1RSB region.  }
\label{fig:q0_vs_mu_rPM}
\end{center}
\end{figure}
In Fig. \ref{fig:q0_vs_mu_rPM}  the value of  $q_0$ is shown at which $b_4(q_0)$ attains its  minimum  as a function of $\xi_4$ for different values of $\gamma$.
In figure \ref{fig:q0_vs_mu_rPM} it is $r=0.5$ but the scenario remains qualitative the same by changing $r$.

\begin{figure}[t!]
\begin{center}
\includegraphics[width=0.99
  \columnwidth]{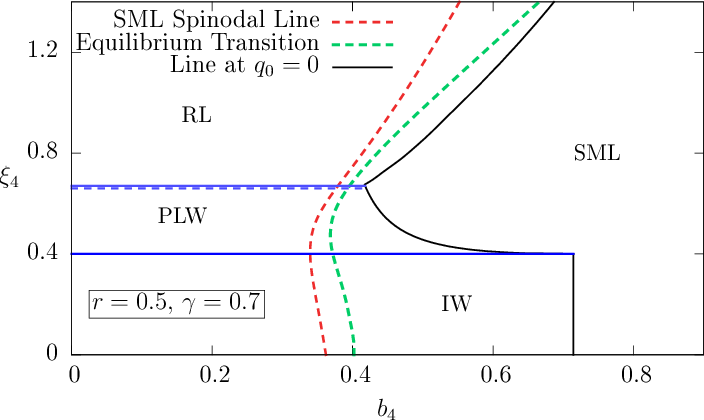}
\caption{ Phase diagram for the model with $r=0.5$ and
  $\gamma=0.7$. For this value of $\gamma$ the spinodal line of the
   RS SML phase enters into the 1RSB region.  
   Dashed red line: spinodal line of the RS SML phase.
   Dashed green line: thermodynamic transition line.
   Solid black line: $b_4(q_0= 0)$ line.  
   Horizontal lines: transition lines between the $m=b=0$ phases.}
\label{fig:b4_r05_G07_allLines}
\end{center}
\end{figure}

\begin{figure}[t!]
\begin{center}
\includegraphics[width=0.99\columnwidth]{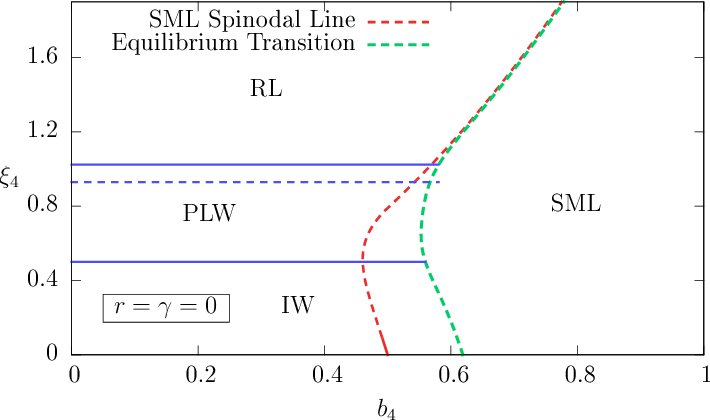}
\caption{ Phase diagram for the model with $r=\gamma=0$.  
   Dashed red line: spinodal line of the RS SML phase.
   Dashed green line: thermodynamic transition line.
   Solid black line: $b_4(q_0= 0)$ line.  
   Horizontal lines: transition lines between the $m=b=0$ phases. 
   }
\label{fig:b4_r0_g0}
\end{center}
\end{figure}

In Fig. \ref{fig:b4_r05_G07_allLines} we report the phase diagram 
for $r=0.5$ and $\gamma=0.7$ in the plane $(b_4,\xi_4)$. 
Similarly, in Fig. \ref{fig:b4_r0_g0} it is shown the phase diagram for
$r=\gamma=0$ when only the 4-body interaction term is present.
In this latter case  as $\xi_4 \to 0$ the transition occurs at $b_4^c = 0.613852$, 
green dashed line in Figure \ref{fig:b4_r0_g0},  in agreement with the result of 
Ref. [\onlinecite{GordonFisherPRL}] 
($b_4 = \gamma_s P_0^2 /(12T)$ in the units of Ref. [\onlinecite{GordonFisherPRL}]).

The unfolding can be also done using the photonics parameters.
The procedure is  conceptually similar but 
more involved because the unfolding equation is now:
\begin{align}
&b  = \frac{1}{2 } \, \mathcal{P} \sqrt{\beta J_0} \,
 \left[ 1 + \left( \frac{\mathcal{P}}{12 \sqrt{\beta J_0}} m^2 -1 \right) \alpha_0 \right]
 m.
\end{align}

Regardless of which approach one uses, this procedure allows to obtain the phase diagram
for general values of the photonic control parameters, 
which eventually may be compared in experimental setups; 
see, in particular, Ref. [\onlinecite{PhysRevLett.111.233903}].

\begin{figure}[t!]
\begin{center}
\includegraphics[width= 0.99 \columnwidth ]{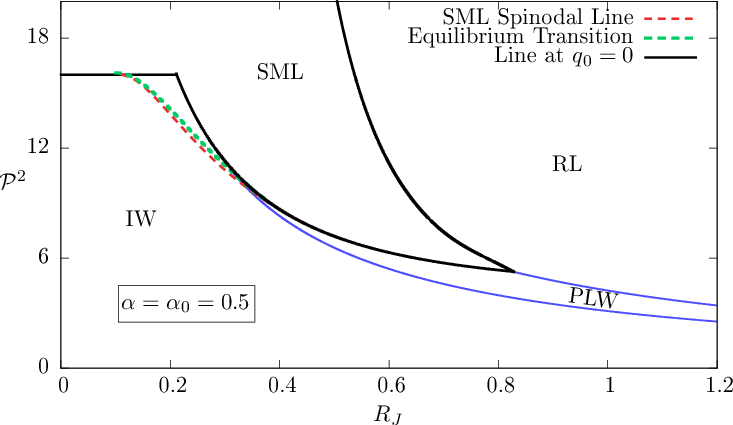}
\caption{ Phase diagram in the photonics parameters for $a = a_0 =
  0.5$.  
    Dashed red line: spinodal line of the RS SML phase.
   Dashed green line: thermodynamic transition line.
   Solid black line: $b_4(q_0= 0)$ line.  
   Blue lines: transition lines between the $m=0$ phases. 
   The transition between the PLW and RL phases
    is continuous with null complexity. }
\label{fig:FM_PM_diagramRL_a0_05_a_05}
\end{center}
\end{figure}

\begin{figure}[t!]
\begin{center}
\includegraphics[width= 0.99 \columnwidth ]{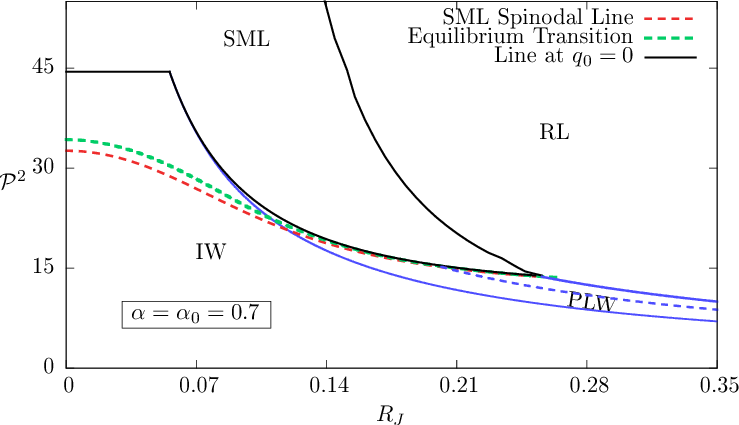}
\caption{ Phase diagram in the photonics parameters for $a = a_0 =
  0.7$.  
   Dashed red line: spinodal line of the RS SML phase.
   Dashed green line: thermodynamic transition line.
   Solid black line: $b_4(q_0= 0)$ line.  
   Blue lines: transition lines between the $m=0$ phases. 
    The dashed blue line is the dynamic
  transition with finite complexity.  }
\label{fig:FM_PM_diagramRL_a0_07_a_07_spinodal}
\end{center}
\end{figure}

\begin{figure}[t!]
\begin{center}
\includegraphics[width= 0.99 \columnwidth ]{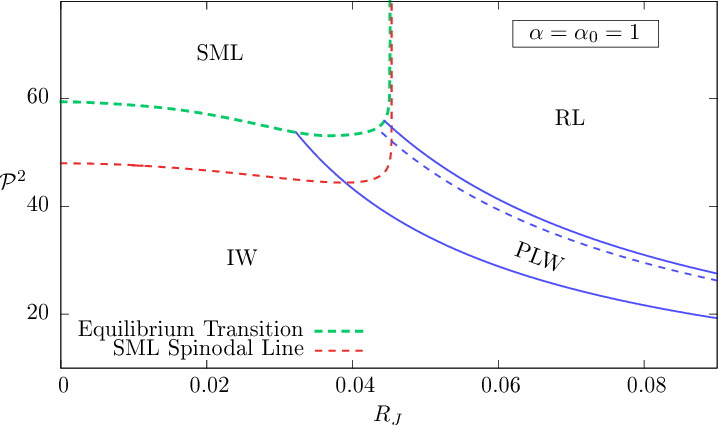}
\caption{ Phase diagram in the photonics parameters for $a = a_0 = 1$.
   Dashed red line: spinodal line of the RS SML phase.
   Dashed green line: thermodynamic transition line.
   Solid black line: $b_4(q_0= 0)$ line.  
   Blue lines: transition lines between the $m=0$ phases. 
    The dashed blue line is the dynamic
  transition with finite complexity. }
\label{fig:FM_PM_diagramRL_a1_a1_spinodal}
\end{center}
\end{figure}

As an example, in 
Figs. \ref{fig:FM_PM_diagramRL_a0_05_a_05}-\ref{fig:FM_PM_diagramRL_a1_a1_spinodal}
we show the phase diagram in the photonic parameters for
$\alpha=\alpha_0=0.5$, $0.7$ and $1$. 
For $\alpha=\alpha_0=1$ only the 4-body interaction term is present, 
corresponding to  the limit of ideally 
closed cavity.  The transition from IW to SML is discontinuous, first order, 
as $\alpha_0$ is large.

\begin{figure}[t!]
\begin{center}
\includegraphics[width= 0.99 \columnwidth ]{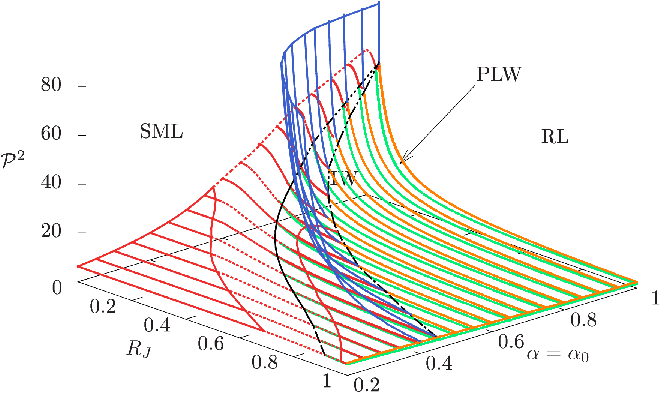}
\caption{ Phase diagram in the photonics parameters for $\alpha =
  \alpha_0$ (and $\beta J_0 = 1$).  The solid (dashed) red lines
  correspond to continuous (discontinuous) IW-SML transition;.
  The blue surface is the RL-SML transition,
  the orange surface the PLW-RL transition
  and the green surface the IW-PLW transition.  
  The two
  black lines mark the intersection  between the
  orange-blue and green-red surfaces, respectively.  }
\label{fig:all_Phase_diagram_a_eq_a0}
\end{center}
\end{figure}

Consider the common experimental situation where we have an
increasing pumping rate $\mathcal{P}$ for fixed degree disorder $R_J$,
$\alpha$ and $\alpha_0$. 
Then:
\begin{itemize}
\item For $R_J$ not too large a direct transition between the IW and the  SML phases 
    is observed as the pumping increases. 
    The transition is robust with respect to the introduction of a small amount of 
disorder;

\item for systems with intermediate disorder, the high pump regime 
remains the ordered SML regime, 
but an intermediate unmagnetized, phase-coherent PLW regime
 appears between SML and IW regime;
\item 
for large $R_J$, a further transition from the SML to the RL
phase is observed at high ${\cal P}$.
Moreover, if $R_J$ exceeds a disorder  threshold the SML disappears and the only high pumping lasing phase remains the RL.
\end{itemize}
This scenario is rather  
general and remains valid for different choices of $\alpha$ and $\alpha_0$.
See for example  Fig. \ref{fig:all_Phase_diagram_a_eq_a0} where the whole phase diagram
for the case $\alpha=\alpha_0$ is reported.  The global picture, however, 
does not change using different values of 
the ratio $\alpha/\alpha_0$.

In general, the value of $\alpha_0$ affects the transition toward the ordered SML
regime: for high $\alpha_0$ the transition is always first order.  In
particular, in the standard closed cavity laser limit
$\alpha_0=1$ and no disorder ($R_J=0$).\cite{GordonFisherPRL} 
On the contrary, if $\alpha_0$ is low, regions in the phase diagram may appear where 
the transition is second order.

The value of $\alpha$ controls the transition to a  
RL regime.
For 
\begin{equation}
\alpha > \alpha_{\text{nl}} =\frac{3-1.76382 \epsilon}{3-1.03703
\epsilon^2}
\end{equation}
 (in this work $\epsilon=1$ and $\alpha_{\text{nl}}= 0.6297\ldots$) the transition is 
toward a RL phase with a 1RSB structure via a RFOT typical of glassy sistems. 
For $\alpha <\alpha_{\rm nl}$ at the lasing threshold the 1RSB phase is unstable, 
and the transition is toward a RL phase with a FRSB structure.  In the latter case the transition
is continuous also in the order parameters.

Though we stress here that in any realistic non-linear 
optical system the $4$-body
interaction is generally expected to be dominant above the lasing threshold, cf. Sect.
\ref{sec:derivation_of_the_model}, in cavity-less
random lasers and in laser cavities with strong leakages 
the contribution from linear interactions plays an important role
on the stimulated emission and 
 the values of $\alpha_0$ and $\alpha$
are not expected to be always close to one.

\begin{figure}[t!]
\begin{center}
\includegraphics[width= 0.99\columnwidth ]{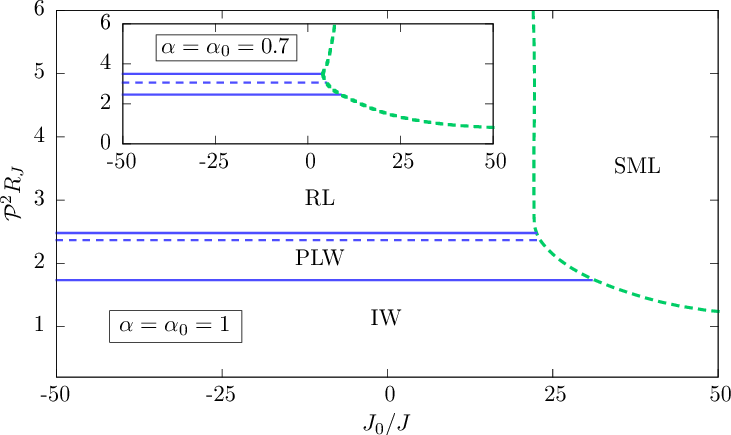}
\caption{ Phase diagram in the parameters $J_0/J$ vs $\mathcal{P}^2
  R_J$ for $\alpha = \alpha_0 = 1$ and $\alpha = \alpha_0 = 0.7$
  (inset).  The green dashed line corresponds to the equilibrium
  thermodynamic transition towards the SML phase, the dashed blue
  line is the dynamic transition between the PLW and RL phases.  
  The continuous blue lines
denote thermodynamic IW/PWL and PWL/RL transitions.}
\label{fig:multiplot_FMPM_digramAP_inverseY}
\end{center}
\end{figure}

A further important point is that the transition from IW
to SML only occurs for a strictly positive value of the coupling
coefficient $J_0$.  This is effectively shown in Fig.
\ref{fig:multiplot_FMPM_digramAP_inverseY}, where the phase diagrams
for $\alpha=\alpha_0=1$ and $\alpha=\alpha_0=0.7$ are displayed in
terms of $R_J^{-1} = J_0/J$ and $\mathcal{P}^2 R_J$.  In standard
passive mode locking lasers, e.g., the coefficient $J_0$ accounts for
the presence of a saturable absorber in the cavity. \cite{HausPaper}
In cavity-less RL this device, or any analogue one, is obviously not
present so that the occurrence of the lasing transition as a
mode-locking is not to be given for granted.  However, as shown in
Fig.  \ref{fig:all_Phase_diagram_a_eq_a0}, starting
from a standard laser supporting passive mode-locking and increasing
the disorder $R_J$, we find that  the IW/SML mode-locking transition acquires the character of a
glassy IW/RL mode-locking transition. This is present even for $J_0<0$, as
explicitly shown in Fig. \ref{fig:multiplot_FMPM_digramAP_inverseY}.

\section{Conclusions}
\label{sec:conclusions}

In this paper we have reported the detailed study of a
statistical mechanical model whose degrees of freedom are complex
continuous spins satisfying a global spherical constraint and
interacting via $2-$ and $4-$body interactions.
The model gives a statistical description of the complex wave amplitude dynamics in
non-linear wave systems. They include, in particular, the multimode laser systems
in presence of a non-perturbative degree of randomness and 
the very interesting limit of random lasers.

We examine in depth the framework of our equilibrium statistical mechanics
description of optical systems in stationary, i.e., off-equilibrium,
non-linear regimes, among which the paradigmatic laser case, and we
discuss the limits of applicability of our theory.

 The dynamics of the electromagnetic modes is described by Langevin
equations with linear and nonlinear terms whose couplings derive from
the spatial overlap of the modes mediated by the spatial profile of
the active medium susceptibility.  We have shown that the presence of
a continuous spectrum of radiative modes leads to a further
contribution to the linear coupling term.  We have critically reported how 
the major difficulty in open and random optically active systems  is
to express the couplings in the slow amplitude mode basis, i.e.,
the modes with a {\em definite} frequency, to which the thermodynamic
approach effectively applies.

The primary condition assumed in the model is the absence of
dispersion in the Master equation, so that the multimode laser  system is
Hamiltonian and the steady-state solution of the associated
Fokker-Plank equation is given by the Gibbs distribution and the
methods of equilibrium statistical mechanics apply.  The purely
dissipative case 
corresponds to some well know
physical situations, like negligible group velocity dispersion and
Kerr effect in standard mode locking lasers and the relevant case of
soliton lasers.  In the general case the condition is hardly
verifiable, though we have shown that in most cases, where the degree
of nonlinearity is  high enough, there are first order transitions
that are expected to be robust by slight
modification of the complex coefficients of the dynamics.

We have reported the results for the mean field theory of the model
that includes the whole complex mode amplitude dynamics when coupled
by both ordered and disordered and linear ($2$-body) and nonlinear
($4$-body) interactions.
The whole phase diagram obtained via the replica theory has been
reported in terms of different sets of external parameters.  We
adopted, in particular, the description typical of glassy systems
undergoing slow relaxation at criticality and dynamic arrest at the
transition. In this case the parameters ($\xi_{2,4}$, $b_{2,4}$) are
the coefficients of the linear and third order terms in the memory
Kernel of a related schematic Mode Coupling Theory
\cite{Crisanti04,Crisanti06, Goetze09} for the hypothetical dynamics
of complex-valued density fluctuations in fluid glass-formers.

Further on, we used an equivalent photonic description in terms of
degree of disorder, source pumping rate and degree of non-linearity.
Four different optical regimes are found: incoherent fluorescence (IW), 
standard mode locking (SML), random lasing (RL) and phase locking wave (PLW). 
In the incoherent
wave the modes oscillate independently as in a paramagnet or in a warm
liquid. In the standard mode locking the modes are coherent both in
phase and in intensity and ultrashort electromagnetic pulses are
generated. This ordered regime corresponds to a ferromagnet, or a
crystal solid.  In the random lasing regime the mode oscillations are
frustrated, so the phases are locked but without global ordering and
any regular pattern in the intensities.  This is a glassy phase,
occurring both as a mean-field window glass, for moderately open cavities
$\alpha>\alpha_{\rm nl}$ or a spin-glass phase, for very open cavities where
linear dumping is not negligible and cannot be treated perturbatively. At last, in the phase locking wave, a
novel regime not foreseen in previous statistical mechanical works,
phase coherence is attained, without any intensity coherence (it is an
unmagnetized phase). This regime takes place between the incoherent
wave and the lasing regimes in presence of a given amount of quenched
disorder and it is achievable only when a model includes the
dynamics of both phases and intensities.

In this context the random lasing threshold is characterized by the
replica symmetry breaking and, at high enough degree of nonlinearity,
a whole region with nonvanishing complexity is observed to anticipate
the transition.  In this sense the light propagating and amplifying in
the disordered medium displays glassy behavior.  We have shown that
this picture is stable against the introduction of a linear coupling
and the transition becomes continuous only when the linear coupling is
actually dominant.  The presence of a small linear coupling, even if not
changing the transition, is shown to alter, nevertheless, the
structure of the phase diagram non-trivially, so that, e. g., even a
transition from standard mode locking to random lasing can be observed
increasing the pumping at fixed degree of disorder.

The inclusion of the full complex amplitudes of the modes as fundamental degrees of freedom,
rather than the mere mode phases as in previous works \cite{AngelaniZamponiPRL,ContiLeuzziPRL09,ContiLeuzziPRB11}
may be prominent
for an experimental test and, in particular, to observe the replica
symmetry breaking predicted by the theory.\cite{Ghofraniha15, Antenucci15}  Indeed,
the measure of the phase-phase correlations required for measuring the
theoretical overlap among the modes is not generally available in the
experiments, as the random lasing emission is hardly intense enough to
use second-harmonic generation techniques, but intensity correlations
can be easily measured.


There are few possible extensions of the previous mean field model
considered in this paper.  To name some, one could consider the
inclusion of correlations in the quenched disordered interactions,
further orders of nonlinear interaction, fluctuations of the spherical
constraint.  Nevertheless, none of these extensions is expected to
change the general mean-field picture described in this work, in
particular the kind of phases and the nature of the transitions
involved.

A more relevant generalization of this approach is the extension to
random laser models beyond the mean-field theory, where new features
may arise and a more direct comparison with experimental setups is
possible, e. g., by generalizing the numerical approach of
Refs. [\onlinecite{Antenucci14b,Antenucci16}] for the pulsed ultrashort mode-locked lasing
in ordered statistical mechanical mode networks to the case of
quenched disordered interactions and random lasing.

Different approaches are, instead, needed to study the evolution of
the system in the general case with a non-negligible presence of
dispersion, to discuss the robustness of the picture presented in this paper.  
The direct analysis of the stochastic dynamics of a finite
number of modes is necessary to this aim.  Moreover, in this case
also the approach to the stationary state would be
accessible, providing an appealing integration of the results
presented here.


\section*{Acknowledgments} The research leading to these
results has received funding from the People Programme (Marie Curie
Actions) of the European Union's Seventh Framework Programme
FP7/2007-2013/ under REA grant agreement no. 290038, NETADIS
project, from the European Research Council through ERC grant
agreement no. 247328 - CriPheRaSy project - and from the Italian MIUR
under the Basic Research Investigation Fund FIRB2008 program, grant
No. RBFR08M3P4, and under the PRIN2010 program, grant code
2010HXAW77-008.


\appendix 
\section{Functional $G[Q,R,m_\sigma, m_\tau]$, 
Eq. \eqref{eq:functional_F_general}.}
\label{averages_quenched_disorder}

In this appendix we give the mains steps to derive the functional 
\eqref{eq:functional_F_general}. The procedure follows the standard derivation
for spherical model, see e.g. Refs. [\onlinecite{Crisanti92}], [\onlinecite{Crisanti06}].
Performing the average in Eq. \eqref{eq:zrep_ave} over the random couplings
of the Hamiltonian (\ref{eq:effective_Hamiltonian_realImmaginary}) with the Gaussian 
probability distribution (\ref{eq:disorder}), leads to
\begin{align}
\label{eq:A1}
 \overline{Z^n} = & \int\, \prod_{j=N}^n\prod_{a=1}^n d\sigma^{\a}_j d\tau^{\a}_j 
  \\
  \nonumber
  &\times \exp \Bigl\{ 
   \frac{2 b_2}{N}  \left[ \left( \sigma^{\a} \right)^2 + \left( \tau^{\a} \right)^2 \right] 
  \\
  &\qquad \quad+  \frac{4 b_4}{ N^3}  \left[ \left( \sigma^{\a} \right)^4 + \left( \tau^{\a} \right)^4 
  + 2 \left( \sigma^{\a} \right)^2 \left( \tau^{\a} \right)^2  \right]
  \Bigr\} 
  \nonumber
   \\
 \nonumber
 &\times \prod_{a,b=1}^n   \exp \Big\{ 
  \frac{ \xi_2}{N}  \left[ \left( \sigma^{\a} \sigma^{\b} \right)^2 + 2 \left( \sigma^{\a} \tau^{\b} \right)^2 + \left( \tau^{\a} \tau^{\b} \right)^2  \right]
  \\
  &
 \qquad\qquad\qquad+ \frac{3 \xi_4}{2 N^3}  \left( F^{\a} F^{\b} \right)^4 
 \Bigr\} \, , 
 \nonumber
\end{align}
where
\begin{align*}
 \left( F^{\a} F^{\b} \right)^4 \equiv & 
 \quad (\sigma^{\a} \sigma^{\b})^4 + (\tau^{\a} \tau^{\b})^4 + 2 (\sigma^{\a} \tau^{\b})^4   \nonumber
 \\
 &  + 4(\sigma^{\a} \sigma^{\b})^2 (\sigma^{\a} \tau^{\b})^2 +  4 (\tau^{\a} \sigma^{\b})^2 (\tau^{\a} \tau^{\b})^2   \nonumber
 \\
 &+ \frac{2}{3} \Bigl[ (\sigma^{\a} \sigma^{\b})^2 (\tau^{\a} \tau^{\b})^2 
 + (\sigma^{\a} \tau^{\b})^2 (\tau^{\a} \sigma^{\b})^2 
 \\
 &+ 4 (\sigma^{\a} \sigma^{\b})(\sigma^{\a} \tau^{\b})(\tau^{\a} \sigma^{\b})(\tau^{\a} \tau^{\b}) \Bigr] \, , \nonumber
\end{align*}
and we have used the shorthand notation
\begin{align}
 \left( \sigma^{a_1} \cdots \sigma^{a_l} \right)^p  \equiv \left( \sum_{j=1}^N \sigma^{a_1}_j \cdots \sigma^{a_l}_j \right)^p \, .
\end{align}
The parameters are those defined in Eqs. (\ref{eq:definition_parameters_b_xi_mu}).

The above expression can be simplified by introducing the matrices $Q_{\a\b}$, $R_{\a\b}$
defined in eqs. \eqref{eq:Q_mat}-\eqref{eq:def_overlap_ReIm}, 
the magnetizations $m_{\sigma;\a}$, $m_{\tau;\a}$ defined in eq. \eqref{eq:def_mag}, 
we have added a replica index to $m_\sigma$ and $m_\tau$ for simplicity of computation,
and the additional matrix
\begin{align}
 T_{\a\b} &= \frac{1}{\mathcal{E}} \sum_{j=1}^N  \Im \left[  a^{a}_j  a^{\b}_j \right]  =  \frac{2}{N} \sum_{j=1}^N \sigma^{\a}_j \tau^{\b}_j. 
\label{eq:def_overlap_T}
\end{align}
This is easily achieved by using the Dirac delta function and the identity
\begin{equation*}
 \int_{-\infty}^{+\infty} dx\, \delta(x -y) = 1,
\end{equation*}
to impose the definitions as, e.g.,
\begin{equation*}
\int _{-\infty}^{\infty} d\Qp_{\a\b}\, \delta \Bigl[\Qp_{\a\b} - 
           [(\sigma^{\a} \sigma^{\b}) + (\tau^{\a} \tau^{\b})]/N \Bigr] = 1,  
\end{equation*}
and similarly for the others.

Using finally the integral representation
\begin{equation*}
  \delta(x-y) = 
  \int_{-\iu \infty}^{i \infty}  \frac{d\hat{x}}{2\pi i}\, \eu^{-\hat{x}\,(x - y)}
\end{equation*}
for the delta function, Eq. \eqref{eq:A1} can be written as
\begin{align}
\label{eq:A4}
  \overline{Z^n} =& \int \mathcal{D}\hat{\Phi} \mathcal{D}\Phi\, 
  \eu^{- N G[\hat{\Phi},\Phi]},
\end{align}
where 
$\Phi = (Q, R, T, \bm{m})$, 
$\bm{m}^a = (m_{\sigma;\a}, m_{\tau;\a})$,
with $\hat{\Phi}$, $\hat{\bm{m}}$ the corresponding hatted quantities,
and the  measure is:
\begin{align*}
\mathcal{D} \hat{\Phi} \mathcal{D} \Phi  &\equiv \mathcal{N}
 \prod_{\a<\b}   d\Qp_{\a\b}\, d\Qm_{\a\b}\,
  \prod_{\a\leq\b}     d\hat{\Qp}_{\a\a}\, d\hat{\Qm}_{\a\a}\,
  \\
 &\quad\times 
 \prod_{\a,\b}   dT_{\a\b}\, d\hat{T}_{\a\b}\,
  \\
 &\quad\times 
  \prod_{\a}  d m_{\sigma;\a}\, d \hat{m}_{\sigma;\a}
  \prod_{\a}  d m_{\tau;\a}\, d \hat{m}_{\tau;\a},
\end{align*}
where $\mathcal{N}$ is an unimportant normalization factor.
The functional in the exponent reads 
\begin{align*}
-G[\hat{\Phi}, \Phi] = 
& \frac{ \xi_2}{2} \sum_{\a\b} \bigl[Q^2_{\a\b} + R^2_{\a\b} + T^2_{\a\b}\bigr] 
\\
&+ \frac{\xi_4}{4} \sum_{\a\b} 
\Bigl[ Q_{\a\b}^4 + R_{\a\b}^4  + 4 Q_{\a\b}^2 R_{\a\b}^2
\nonumber
\\
&
\quad\qquad + \frac{3}{4}T_{\a\b}^4  
+ 3 T_{\a\b}^2\bigl( Q_{\a\b}^2 + R_{\a\b}^2\bigr)
\\
&\quad\qquad
+ \frac{1}{4}T_{\a\b}^2T_{\b\a}^2
  + T_{\a\b} T_{\b\a} \bigl( Q_{\a\b}^2 - R_{\a\b}^2\bigr)
   \Bigr] 
  \\
&
\quad
+ \sum_{\a} k( m_{\sigma;\a}, m_{\tau;\a})
  \\
  & 
  - \frac{1}{2}
    \sum_{\a\b} \bigl[ \hat{Q}_{\a\b} Q_{\a\b} + \hat{R}_{\a\b} R_{\a\b} +  \hat{T}_{\a\b} T_{\a\b} \bigr]
  \\
  &
  \nonumber
  - \frac{1}{\sqrt{2}} \sum_a 
    \hat{\bm{m}}^a\cdot \bm{m}^a
    + \ln Z[\hat{\Phi}],
\end{align*}
where $k(x,y)$ is given in Eq. \eqref{eq:definition_k},
 the 'dot' denotes standard scalar product, and
\begin{align}
\label{eq:A5}
Z&[\hat{\Phi}] =  \int \prod_{\a=1}^n d \sigma_{\a} d\tau_{\a} 
 \\
 &
\times
  \exp  \left[ 
 \frac{1}{2} \bigl(
  \sigma \hat{A} \sigma + \tau\hat{B} \tau  + 2 \sigma\hat{T}\tau
   \bigr)
 -\hat{m}_{\sigma}\sigma - \hat{m}_{\tau}\tau
 \right],
 \nonumber 
\end{align}
with summation over replica indexes assumed, and 
$\hat{A}_{\a\b} = \hat{Q}_{\a\b} + \hat{R}_{\a\b}$ while
$\hat{B}_{\a\b} = \hat{Q}_{\a\b} - \hat{R}_{\a\b}$. 
Note that as usual an extra factor $2$ is added when needed to the diagonal elements 
of matrices to account for the extra $1/2$ factor coming from the symmetry of the matrix.

In the thermodynamic limit $N\to\infty$ the integrals in \eqref{eq:A4} 
are dominated by the largest value of the exponent and hence they can be evaluated 
using the saddle point  method. Note that  the order of the $N \to \infty$ and  $n \to 0$ limits is 
exchanged. 

Stationarity of $G[\hat{\Phi},\Phi]$ with respect to variations of $T$ and $\hat{T}$ leads 
respectively to the 
saddle point equations
\begin{equation}
\label{eq:A6}
\begin{split}
 \xi_2 T_{\a\b} + &\frac{\xi_4}{4}\bigl[
           3T_{\a\b}^3 + 6T_{\a\b} (Q_{\a\b}^2 + R_{\a\b}^2)\\
           &+T_{\a\b}T_{\b\a}^2 
           + 2T_{\b\a} (Q_{\a\b}^2 - R_{\a\b}^2)
                          \bigr] = \frac{1}{2}\hat{T}_{\a\b},
  \end{split}                        
\end{equation}
\begin{equation}
\label{eq:A7}
  T_{\a\b} = 2\langle\sigma^a\tau^b\rangle,
\end{equation}
where the average is performed with the probability measure defined in \eqref{eq:A5}.
The original Hamiltonian \eqref{eq:effective_Hamiltonian_3} is invariant under a 
global phase rotation $a \to a \, \eu^{\iu \varphi}$, and hence $T_{\a\b}$ must be symmetric. 
Thus the only possible solution to \eqref{eq:A7} is  
$T_{\a\b} = 2 \langle\sigma^\a\rangle \langle\tau^\b\rangle = 2 m_{\sigma}m_{\tau}$
because  without an explicit breaking of the replica equivalence, 
{\sl vectorial} RSB, all replicas are identical and single replica  quantities cannot depend on
the replica. 
This in turns implies that also $\hat{T}_{\a\b}$ cannot depend on $\a$ and $\b$.
The only possible solution to \eqref{eq:A6} with replica independent $\hat{T}_{\a\b}$ and 
$T_{\a\b}$ is $\hat{T}_{\a\b} = T_{\a\b} = 0$.
Then $m_\sigma$ and $m_\tau$ cannot be different from zero simultaneously.

The other consequence of $\hat{T}_{\a\b} = T_{\a\b} = 0$ is that integral over
$\sigma$ and $\tau$ in \eqref{eq:A5} decouples and hence 
$G[\hat{\Phi}, \Phi]$ for the solution  $\hat{T}_{\a\b} = T_{\a\b} = 0$ can be written as
\begin{equation}
\label{eq:A8}
\begin{split}
  -G[\hat{\Phi}, \Phi] &= F[Q,R,\bm{m}]  
  - \frac{1}{4}
    \bigl[ \hat{A} A+ \hat{B} B \bigr] \\
 &     - \frac{1}{\sqrt{2}}
    \hat{\bm{m}}\cdot \bm{m}
    + W_0[\hat{m}_\sigma, \hat{A}]  + W_0[\hat{m}_\tau, \hat{B}].
 \end{split}   
\end{equation}
where again the sum over replicas indexes is assumed,
\begin{equation}
 F[Q,R,\bm{m}]  = \frac{1}{2} \sum_{\a\b} g\bigr(Q_{\a\b}, R_{\a,b}\bigr)
 + \sum_{\a} k( m_{\sigma;\a}, m_{\tau;\a}),
\end{equation}
with $g(x,y)$ given in \eqref{eq:definition_g},
and $W_0[J,K]$ is the massless limit $\Delta\to 0$ of the connected generating functional
of the gaussian theory
\begin{equation}
  W_\Delta[J,K] = \ln \int d\sigma \exp\Bigl[ -\frac{1}{2}\sigma\Delta\sigma  + \frac{1}{2}\sigma K \sigma
                  + J\sigma\bigr].
\end{equation}
The transformation $(Q,R) \to (A,B)$ is orthogonal and hence the form
$\hat{Q} Q + \hat{R} R$ is invariant, but gain an additional $1/2$ factor.
Even if $m_\sigma$ and $m_\tau$ cannot be simultaneously non null, 
we keep both to treat the two cases at the same time. Also we maintain 
their replica index when needed to simplify the notation.

Stationarity  with respect to $\hat{Q}$ and $\hat{R}$ is equivalent to stationarity 
with respect to $\hat{A}$ and $\hat{B}$.
The stationarity of $G$ with respect to $\hat{m}_\sigma$ and $\hat{A}$ leads to the saddle point 
equations
\begin{align}
 \frac{\partial }{\partial \hat{m}_\sigma}  W_0[\hat{m}_\sigma, \hat{A}] &=  
               \frac{1}{\sqrt{2}} m_\sigma, \\
 \frac{\partial }{\partial \hat{A}}  W_0[\hat{m}_\sigma, \hat{A}] &=  
             \frac{1}{4} A,
\end{align}
from which one recognizes in \eqref{eq:A8}
the double Legendre transform of $W_0[\hat{m}_\sigma, \hat{A}]$.

The double Legendre transform $\Gamma_{\Delta}[G,\varphi]$  of $W_\Delta[J,K]$ 
is defined as
\begin{equation}
\begin{split}
   \Gamma_{\Delta}[G,\varphi] + W_\Delta[J,K] &= J\varphi + \frac{1}{2}\varphi K \varphi 
                   + \frac{1}{2} G K, \\
                   \frac{\partial }{\partial J}  W_\Delta[J, K] &=  \varphi, \\
    \frac{\partial }{\partial K}  W_\Delta[J, K] &=  \frac{1}{2}\bigl[G + \varphi \varphi\bigr].
 \end{split}                  
\end{equation}
Then comparison with \eqref{eq:A8} shows that 
\begin{equation}
 -W_0[\hat{m}_\sigma, \hat{A}] +  \frac{1}{4} \hat{A} A + \frac{1}{\sqrt{2}} \hat{m}_\sigma m_\sigma 
 = \Gamma_0[G,\varphi],
\end{equation}
with
\begin{equation}
  \varphi = \frac{1}{\sqrt{2}} m_\sigma, 
  \qquad
  \frac{1}{2} A = G + \varphi \varphi.
\end{equation}
The double Legendre transform $\Gamma_\Delta[G,\varphi]$ for the Gaussian (free) theory
is equal to,\cite{DeDomMar64a,CorJacTom74,Haymaker91} 
\begin{equation}
  \Gamma_\Delta[G,\varphi] = \frac{1}{2}\varphi\Delta\varphi + \frac{1}{2}\Delta G 
        +\frac{1}{2}\Tr\ln G^{-1},
\end{equation}
and can be derived from general arguments or by evaluating it directly. 
Then, neglecting unnecessary constants, 
\begin{align}
 -W_0[\hat{m}_\sigma, \hat{A}] &+  \frac{1}{4} \hat{A} A 
      + \frac{1}{\sqrt{2}} \hat{m}_\sigma m_\sigma =
      \nonumber
      \\
      &=-\frac{1}{2}\Tr\ln\bigl[A - m_\sigma\otimes m_\sigma\bigr]\\
      &=-\frac{1}{2}\Tr\ln A + \frac{1}{2}  m_\sigma A^{-1} m_\sigma+ O(n^2).
      \nonumber
 \end{align}
Stationarity of $G$ with respect to $\hat{m}_\tau$ and $\hat{B}$ leads to a similar result.
Then collecting all terms 
\begin{equation}
\label{eq:A18}
 \begin{split}
 -G[\Qp,\Qm,&m_\sigma,m_\tau] = 
          \frac{1}{2} \sum_{a,b}  g (\Qp_{\a\b},\Qm_{\a\b})  
         \\
 &+  n\,k(m_{\sigma}, m_{\tau}) 
 \\
 &+ \frac{1}{2} \Tr\ln A   
  - \frac{m_{\sigma}^2}{2} \sum_{a,b} (A^{-1})_{\a\b}
 \\
& + \frac{1}{2} \Tr\ln B -  \frac{m_{\tau}^2}{2} \sum_{a,b}   (B^{-1})_{\a\b},
\end{split}
\end{equation}
which completes the derivation of  \eqref{eq:functional_F_general}:

\section{Solutions of the stationary equations}
\label{App:Solutions}

The equations for the order parameters $Q_{\a\b}$, $R_{\a\b}$ and
magnetizations $m_\sigma$, $m_\tau$ follow from stationarity
of the functional $G[Q,R,m_\sigma, m_\tau]$.

Introducing the function $\Lambda(x,y)$, defined in \eqref{eq:Lambda},
and noticing that
$(\partial/\partial y) g(x,y) = \Lambda(y,x)$
because of  the symmetry $g(y,x) = g(x,y)$,
stationarity with respect to variations of $Q_{\a\b}$ and $R_{\a\b}$ with $\a\not=\b$
leads to the 
saddle point equation:
\begin{equation}
\label{eq:B1}
\begin{split}
    \Lambda (Q_{\a\b},R_{\a\b})& + (A^{-1})_{\a\b} 
    + m_\sigma^2 \Bigl[\sum_{\b} (A^{-1})_{\a\b}\Bigr]^2
\\
&+  (B^{-1})_{\a\b}
  +   m_\tau^2 \Bigl[\sum_{\b} (B^{-1})_{\a\b}\Bigr]^2
 = 0,
\end{split}
\end{equation}
\begin{equation}
\label{eq:B2}
\begin{split}
    \Lambda (R_{\a\b},Q_{\a\b})& + (A^{-1})_{\a\b} 
    + m_\sigma^2 \Bigl[\sum_{\b} (A^{-1})_{\a\b}\Bigr]^2
\\
&-  (B^{-1})_{\a\b}
  -   m_\tau^2 \Bigl[\sum_{\b} (B^{-1})_{\a\b}\Bigr]^2
 = 0.
\end{split}
\end{equation}
To obtain these equations we have used the property that the sum off all elements of a row
(column) of overlap matrices does not depended on the row (column) index.
 
The function $g(x,y)$ is even in $x$ and $y$, 
see \eqref{eq:definition_g},  then $\Lambda(-x,y) = -\Lambda(x,y)$
while $\Lambda(x,-y) = \Lambda(x,y)$.
Consequence of this, if $m_\tau = 0$ then the solution to \eqref{eq:B1}-$\eqref{eq:B2}$
is $R_{\a\b} = Q_{\a\b}$ for $\a\not=\b$.
Vice versa, if $m_\sigma = 0$ then the solution is $R_{\a\b} = -Q_{\a\b}$ for $\a\not=\b$.
If $m_\sigma = m_\tau = 0$ then both solutions exist, and are degenerate. 
Note that 
$R_{\a\b} = Q_{\a\b}$  implies  $\langle\tau^\a\tau^\b\rangle =0$
while $R_{\a\b} = -Q_{\a\b}$  that $\langle\sigma^\a\sigma^\b\rangle =0$.

For what concerns the diagonal elements $Q_{\a\a}$ they are all equal to $1$ because of the 
spherical constraint \eqref{eq:spherical_constraint_1}.
Similarly $R_{\a\a}=\overline{R}$ for all $a$ with  $\overline{R}$ determined   by stationarity
via \eqref{eq:B2} evaluated for $\a=\b$.

The analysis of stationarity of $G[Q,R,m_\sigma, m_\tau]$ 
with respect to variations of $m_\sigma$ or $m_\tau$ simplifies 
if the solution is specified. The general treatment does not add anything more. 
Consider the solution with $m_\tau = 0$,  the case $m_\sigma=0$ is similar. Then
stationarity leads to the saddle point equation:
\begin{equation}
\label{eq:B3}
  n\, \frac{\partial}{\partial m_\sigma} k(m_\sigma,0) - m_\sigma \sum_{\a\b} (A^{-1})_{\a\b} = 0,
\end{equation}
which, making use of the identity
\begin{equation}
 \frac{1}{n} \sum_{\a\b} A_{\a\b} = \frac{1}{\displaystyle  \frac{1}{n} \sum_{\a\b} (A^{-1})_{\a\b}} 
\end{equation}
gives
\begin{equation}
\label{eq:B5}
  m_\sigma =  \left[\frac{1}{n} \sum_{\a\b} A_{\a\b}\right]
                          \frac{\partial}{\partial m_\sigma} k(m_\sigma,0).
\end{equation}
Inserting this form into \eqref{eq:A18}, the functional $G$ can be written in the equivalent form
\begin{equation}
 \begin{split}
 -G[\Qp,\Qm,&m_\sigma, 0] = 
          \frac{1}{2} \sum_{a,b}  g (\Qp_{\a\b},\Qm_{\a\b})  
 \\
 &+ \frac{1}{2} \Tr\ln A   
  - \frac{b^2}{2} \sum_{a,b} A_{\a\b}
 \\
& + \frac{1}{2} \Tr\ln B +  n\,k(m_{\sigma}, 0)
\end{split}
\end{equation}
where $b = (\partial/\partial m_\sigma)k(m_\sigma,0)$. 
This form, besides the last term, is the same one would get for a model described by  
the Hamiltonian \eqref{eq:effective_Hamiltonian_realImmaginary}, but
random couplings with $J_0^{(p)}=0$, and a uniform external field $\sqrt{2} b/\beta$ 
coupled with the  $\sigma_j$.\cite{CrisantiLeuzzi_sp_NPB}
The factor $\sqrt{2}$ follows from the definition \eqref{eq:def_mag} of $m_\sigma$ and ensures
 that 
\begin{equation}
\label{eq:B7}
  m_\sigma =  \frac{b}{n} \sum_{\a\b} A_{\a\b},
\end{equation}
so that \eqref{eq:B5} is satisfied.

For overlap matrices with the Parisi replica symmetry breaking scheme\cite{MPVBook} 
the equations can be written more explicitly in terms of their Replica Fourier 
Transform,\cite{ParisiFourier,ReplicaFourier97, CrisantiRFT} 
defined for a generic matrix $M_{\a\b}$ in eqs. (\ref{eq:def_RFT}).

The functional $G[Q,R,m_\sigma, m_\tau]$ then takes the form \eqref{eq:freeEnergy_RSB}, 
see eg. [\onlinecite{CrisantiLeuzzi_sp_NPB}], while the saddle point equations 
\eqref{eq:B1} becomes, as $p_0=n\to 0$:
\begin{equation}
 \Lambda(Q_0,R_0) - \frac{A_0 - m_\sigma^2}{(A_{\widehat{1}})^2} 
         - \frac{B_0 - m_\tau^2}{(B_{\widehat{1}})^2} = 0,
\end{equation}
and 
\begin{equation}
\begin{split}
\Lambda(Q_r,R_r) - &\Lambda(Q_{r-1},R_{r-1}) \\
 & 
 - \frac{A_r - A_{r-1}}{A_{\widehat{r}} A_{\widehat{r+1}} }
 - \frac{B_r - B_{r-1}}{B_{\widehat{r}} B_{\widehat{r+1}} } = 0,
 \end{split}
\end{equation}
for $r=1, \ldots, \mathcal{R}$.
Similarly  \eqref{eq:B2} reads:
\begin{equation}
 \Lambda(R_0,Q_0) - \frac{A_0 - m_\sigma^2}{(A_{\widehat{1}})^2} 
                       + \frac{B_0 - m_\tau^2}{(B_{\widehat{1}})^2} = 0,
\end{equation}
and, for $r=1, \ldots, \mathcal{R}+1$,
\begin{equation}
\begin{split}
\Lambda(R_r,Q_r) - &\Lambda(R_{r-1},Q_{r-1}) \\
 & 
 - \frac{A_r - A_{r-1}}{A_{\widehat{r}} A_{\widehat{r+1}} }
 + \frac{B_r - B_{r-1}}{B_{\widehat{r}} B_{\widehat{r+1}} } = 0,
 \end{split}
\end{equation}
with the  boundary value $Q_{\mathcal{R}+1} = 1$ and 
$R_{\mathcal{R}+1} = \overline{R}$.
Finally for the case $m_\tau = 0$ \eqref{eq:B5}, or equivalently \eqref{eq:B7}, 
simply becomes
\begin{equation}
  m_\sigma = b\, A_{\hat{1}}.
\end{equation}

The parameters $p_r$, except for $p_0=n$ and $p_{\mathcal{R}+1} = 1$ are not determined.
If we require that $G$ be stationary  with respect also to variations of $p_r$, then we have the 
additional set of equations
\begin{align*}
  g(Q_r,\, & R_r)  - g(Q_{r-1},R_{r-1}) =
 \\
 &
    \frac{1}{p_r^2} \log \frac{A_{\widehat{r}}}{A_{\widehat{r+1}}}-\frac{A_r-A_{r-1} }{p_r A_{\widehat{r}}}
 \\
 &
  +(A_r-A_{r-1}) \left[ \sum_{t=1}^{r-1} \frac{A_t-A_{t-1}}{A_{\widehat{t}} A_{\widehat{t+1}}} +\frac{A_0-m_\sigma^2}{(A_{\widehat{1}})^2} \right] 
 \\
 & +  \frac{1}{p_r^2} \log \frac{B_{\widehat{r}}}{B_{\widehat{r+1}}}
 - \frac{B_r-B_{r-1}}{p_r B_{\widehat{r}}} 
 \\
 &+ (B_r-B_{r-1}) \left[ \sum_{t=1}^{r-1} \frac{B_t-B_{t-1}}{B_{\widehat{t}} B_{\widehat{t+1}}} +\frac{B_0-m_\tau^2}{(B_{\widehat{1}})^2} \right] 
 = 0, 
 \\
 &
 \qquad 
 r=1, \ldots \mathcal{R}
 \, .
\end{align*}
for $r=1, \ldots, \mathcal{R}$ which leads to the {\sl static} solution.
The other case of interest, called the {\sl dynamical} solution,  
is the request that the state be {\sl marginally stable}, that is that the relevant eigenvalue of
fluctuations about the stationary point vanishes. For example, this request the 1RSB solution 
leads to the additional equation \eqref{eq:x_dyn}. See also Appendix \ref{App:Fluct}.

\section{Saddle Point Stability Analysis}
\label{App:Fluct}

The evaluation of the integrals in \eqref{eq:A4} by the saddle point method requires that 
at the stationary point the functional $G[\hat{\Phi},\Phi]$ attains its minimum value.  Convexity 
of the Legendre transform ensures that $G[Q,R,m_\sigma,m_\tau]$ should also be minimum at
the saddle point.
However as $n<1$ the space dimension of $Q_{\a\b}$ and $R_{\a\b}$ with $\a\not=\b$ 
becomes negative turning the minimum into a maximum. 
Thus as $n\to 0$ the functional $G[Q,R,m_\sigma,m_\tau]$ evaluated at the saddle point 
gives the maximum value with respect to variations of $Q$ and $R$
and the minimum value for variations in $m$.

The second observation is that  $(\delta Q, \delta R)$ and $\delta m$ fluctuations can be studied separately. 
Cross-correlations $\delta Q \delta m$
and $\delta R\delta m$ originate only from  the $m_\sigma^2\sum A^{-1}$, 
or $m_\tau^2\sum B^{-1}$, term,
see \eqref{eq:A18}.
The quadratic form $\delta^2G$ controlling  the fluctuations 
about the saddle point can then be studied separately 
in the two orthogonal spaces $(\delta Q, \delta R, 0)$ and $(0,0,\delta m)$.
Stated differently, this means that  $m$ fluctuations can be studied for fixed $(Q,R)$ and
$m$ fluctuations for fixed $(Q,R)$.

The stability analysis of $m$ fluctuations 
 it is not dissimilar to that of an
ordinary mean-field ferromagnet. 
Fixing $Q$ and $R$, from \eqref{eq:A18}, and using  \eqref{eq:B3}, 
the $m_\sigma$ fluctuations in the $m_\tau=0$ solution are stable if
\begin{equation}
\begin{split}
- \frac{\partial^2}{\partial m_\sigma^2}& k(m_\sigma,0) + \frac{1}{n}\sum_{\a\b} (A^{-1})_{\a\b}
 = \\
 & = \left[
 - \frac{\partial^2}{\partial m_\sigma^2} + \frac{1}{m}  \frac{\partial}{\partial m_\sigma}
 \right] k(m_\sigma,0)  \geq 0.
 \end{split}
\end{equation}
For the model discussed in main text
this condition becomes, see \eqref{eq:definition_k},
\begin{equation}  
  8 b_4 m_\sigma^2 \geq 0,
\end{equation}
and hence for non-negative $b_4$ the solution $m_\sigma\not=0$, if it exists, it is always stable.

The analysis of $(Q,R)$ fluctuations  is more complex and depends on the RSB structure.
To keep the discussion as simple as possible we shall limit ourself to the case 
$m_\sigma=m_\tau = 0$. 
The study of the eigenvalues of the hessian matrix $M^{\a\b;\c\d}$ of the $(Q,R)$ 
fluctuations about the 
saddle point is unaffected by this choice, however it simplifies the calculations.
Moreover,  the main consequence of not null $m_\sigma$, or $m_\tau$,
is a non vanishing  $Q_0$ and $R_0$. This modifies the expression of some eigenvalues
but not that of the  relevant eigenvalues controlling the stability of the saddle point.
The number of different eigenvalues, and their expression, depends on the
RSB structure. We shall limit ourself to the RS and 1RSB cases only.

The eigenvalues of the Hessian matrix of the $(Q,R)$ fluctuations 
are  solution of the eigenvalue equation
\begin{align}
\label{eq:eig_eq_Q}
 \frac{1}{2} \sum_{\c\d} \Bigl[
                            \lsb{QQ}M^{\a\b;\c\d}\delta Q^{\c\d} + 
                            \lsb{QR}M^{\a\b;\c\d}\delta R^{\c\d}
                            \Bigl] &= \Lambda\, \delta Q^{\a\b}
                            \\
\label{eq:eig_eq_R}
 \frac{1}{2} \sum_{\c\d} \Bigl[
                            \lsb{RQ}M^{\a\b;\c\d}\delta Q^{\c\d} + 
                            \lsb{RR}M^{\a\b;\c\d}\delta R^{\c\d}
                            \Bigl] &= \Lambda\, \delta R^{\a\b} 
\end{align}
where
\begin{align}
 \lsb{QQ}M^{\a\b;\c\d} &=  \frac{\partial^2 G}{\partial Q_{\a\b} \partial Q_{\c\d}} 
 \\
                                      & = -\Lambda'(Q_{\a\b},R_{\a\b})\, \delta_{\a\b,\c\d}
                                                 + \lsb{+}M^{\a\b;\c\d},
\nonumber
\\                                                 
 \lsb{QR}M^{\a\b;\c\d} &=  \frac{\partial^2 G}{\partial Q_{\a\b} \partial R_{\c\d}} 
 \\
                                      & = -\dot{\Lambda}(R_{\a\b},Q_{\a\b})\, \delta_{\a\b,\c\d}
                                                 + \lsb{-}M^{\a\b;\c\d},
\nonumber                                                 
\end{align}
\begin{align}
 \lsb{RR}M^{\a\b;\c\d} &=  \frac{\partial^2 G}{\partial R_{\a\b} \partial R_{\c\d}} 
 \\
                                      & = -\Lambda'(R_{\a\b},Q_{\a\b})\, \delta_{\a\b,\c\d}
                                                 + \lsb{+}M^{\a\b;\c\d},
\nonumber
\\                                                 
 \lsb{RQ}M^{\a\b;\c\d} &=  \frac{\partial^2 G}{\partial R_{\a\b} \partial Q_{\c\d}} 
 \\
                                      & = -\dot{\Lambda}(Q_{\a\b},R_{\a\b})\, \delta_{\a\b,\c\d}
                                                 + \lsb{-}M^{\a\b;\c\d},
\nonumber                                                 
\end{align}    
and
\begin{equation}
\begin{split}
  \lsb{\pm}M^{\a\b;\c\d} &= \bigl[ (A^{-1})_{\a\c} (A^{-1})_{\b\d} +  
                                                      (A^{-1})_{\a\d} (A^{-1})_{\b\c} \bigr]
\\                                                      
	              &\ \pm           \bigl[ (B^{-1})_{\a\c} (B^{-1})_{\b\d} +  
                                                      (B^{-1})_{\a\d} (B^{-1})_{\b\c} \bigr].                                                   
\end{split}                                                      
\end{equation}
The function $\Lambda(x,y)$ is defined in \eqref{eq:Lambda},                                   
and we have denoted derivatives with respect to $x$  by a prime 
and derivatives with respect to $y$ by an overdot.
With this notation $\Lambda(x,y) = g'(x,y)$.
The factor $1/2$ in the eigenvalue equations follows from the symmetry 
$M^{\a\b;\d\c} = M^{\a\b;\c\d}$ for $\c\not=\d$, with the implicit assumption 
that the diagonal element $M^{\a\b;\c\c}$ is multiplied by $2$. 
The ``vectors'' $\delta Q^{\a\b}$ and $\delta R^{\a\b}$, also symmetric under index exchange,
represent the fluctuation of $Q$ and $R$ from the saddle point value. Note that
$\delta Q^{\a\a} = 0$ because its value $Q_{\a\a} = 1$ is fixed by
the  spherical constraint \eqref{eq:spherical_constraint_1}, and not from stationarity,
and cannot fluctuate.

The eigenvalue equations \eqref{eq:eig_eq_Q}-\eqref{eq:eig_eq_R} can be  solved
easily
for the RS solution, especially  if $Q_0=R_0=0$. For the 1RSB solution 
the calculation is still feasible using  the decomposition of Ref. [\onlinecite{Crisanti92}],
but becomes rather cumbersome.  
The more systematic construction of Ref.  [\onlinecite{DeDominicis}] 
becomes also rapidly involved as the number of RSB steps increases.
The simplest approach is to solve the eigenvalue equations in the Fourier Replica Space 
using the Replica Fourier  Transform.\cite{ReplicaFourier97, CrisantiRFT}

In the Replica space the metric is defined by the overlap or (co)-distance 
$\a\cap\b = r$ between two replicas. This means that replica $\a$ and $\b$  belong to the same
Parisi box of size $p_r$ but to two distinct boxes of size $p_{r+1}$. 
This metric defines  an ultrametric geometry and hence the
four replicas  $\a,\b,\c,\d$ of $M^{\a\b;\c\d}$
can be found only in two different type of configurations.

When the direct overlaps $\a\cap\b=r$ and $\c\cap\d = s$ are different 
$M^{\a\b;\c\d}$ can depend on only one cross-overlap:
\begin{equation}
 M^{\a\b;\c\d} = M^{r,s}_t,
 \ \ t = \max\bigl( \a\cap\c, \a\cap\d, \b\cap\c, \b\cap\d\bigr).
\end{equation}
These configurations are called {\sl Longitudinal-Anomalous} (LA).
 When $r=s$ two types of configurations are possible. The first are the LA configurations
 with $r=s$. However a second, different type, of configurations appears 
when the cross-overlap between $(\a,\b)$ and $(\c,\d)$ is larger than $r$. For these
configurations $M^{\a\b;\c\d}$ depends on  two cross-overlaps:
\begin{equation}
\begin{split}
  M^{\a\b;\c\d} = M^{r,r}_{u,v},
   \qquad
        \begin{split}
                u &= \max(a\cap c, a\cap d),\\
                v &= \max(b\cap c, b\cap d),
         \end{split}
         \end{split}
\end{equation}
with $u,v \geq r+1$. These configurations are called  {\sl Replicon}.\footnote{ 
Boundary terms from LA configurations contribute to $M^{r,r}_{u,v}$. 
These, however, are projected out by the Replica Fourier Transform and we can safely ignore 
them. For a more detailed discussion we refer to Ref. [\onlinecite{CrisantiRFT}].}
The Replicon and LA sectors are orthogonal to each other 
and the eigenvalue equations in the two sectors 
decoupled thus block diagonalizing the Hessian matrix.

In the Replicon sector the eigenvalue equations  in the Fourier Replica Space read:
 \begin{align}
 \label{eq:Rep_Q}
 \lsb{QQ}M^{r,r}_{\hat{k},\hat{l}}\, \delta Q^{r,r}_{\hat{k},\hat{l}} +
 \lsb{QR}M^{r,r}_{\hat{k},\hat{l}}\, \delta R^{r,r}_{\hat{k},\hat{l}} 
  &= \Lambda\,  \delta Q^{r,r}_{\hat{k},\hat{l}} , 
  \\
 \label{eq:Rep_R}
 \lsb{RQ}M^{r,r}_{\hat{k},\hat{l}}\, \delta Q^{r,r}_{\hat{k},\hat{l}} +
 \lsb{RR}M^{r,r}_{\hat{k},\hat{l}}\, \delta R^{r,r}_{\hat{k},\hat{l}} 
  &= \Lambda\,  \delta R^{r,r}_{\hat{k},\hat{l}} , \
  \end{align}
where  
\begin{eqnarray}
 \lsb{QQ}M^{r,r}_{\hat{k},\hat{l}} &=& -\Lambda'(Q_r, R_r) 
                                                            + \frac{1}{A_{\hat{k}}\,A_{\hat{l}}}
						    + \frac{1}{B_{\hat{k}}\,B_{\hat{l}}}
						    \\
 \lsb{RR}M^{r,r}_{\hat{k},\hat{l}} &=& -\Lambda'(R_r, Q_r) 
                                                            + \frac{1}{A_{\hat{k}}\,A_{\hat{l}}}
						    + \frac{1}{B_{\hat{k}}\,B_{\hat{l}}}
						    \\
 \lsb{RQ}M^{r,r}_{\hat{k},\hat{l}} &=& -\dot{\Lambda}(Q_r, R_r) 
                                                            + \frac{1}{A_{\hat{k}}\,A_{\hat{l}}}
						    - \frac{1}{B_{\hat{k}}\,B_{\hat{l}}}
						    \\
 \lsb{QR}M^{r,r}_{\hat{k},\hat{l}} &=& \lsb{RQ}M^{r,r}_{\hat{k},\hat{l}}.
\end{eqnarray}
For each $(r,k,l)$, with $r=0,\dotsc,\mathcal{R}$  and 
and $k,l  = r+1, \dotsc, \mathcal{R}+1$,
the Replicon eigenvalues 
$\lsb{1,2}\Lambda(r;k,l)$ 
are  obtained from the eigenvalue of a $2\times 2$ real
symmetric matrix, and are  hence real. Each eigenvalue has multiplicity:
\begin{equation}
 \mu(r; k,l) = \frac{p_0}{2}\, \overline{\delta}_k\, \overline{\delta}_l\, \delta_r(k,l)
\end{equation}
where for $k> r+1$
\begin{equation}
\label{eq:delta_bar}
 \overline{\delta}_k = \frac{1}{p_k} - \frac{1}{p_{k-1}},
\end{equation}
 while  $\overline{\delta}_{r+1}= 1/p_{r+1}$ for $k=r+1$, and
\begin{equation}
 \delta_r(k,l) = p_r - (1 + \delta_{k,r+1} + \delta_{l,r+1})\, p_{r+1}.
\end{equation}

In the LA sector the eigenvalue equations are more complex. 
The LA eigenvalues $\Lambda(k)$ are labelled by the single index 
$k=0,\dotsc, \mathcal{R}+1$ and 
for each $k$ are obtained from the eigenvalues of a 
$(2\mathcal{R}+3)\times(2\mathcal{R}+3)$ matrix.

In the Fourier Replica Space the eigenvalue equations in the LA sector take the form:
\begin{align}
\label{eq:LAQ}
& \sum_{s=0}^{\mathcal{R}+1} \Bigl[
  \bigl[\delta_{r,s}\, \lsb{QQ}M^{r,r}_{\widehat{r+1},\hat{k}} 
             + \frac{1}{4}\delta^{(k-1)}_s\, \lsb{QQ}M^{r,s}_{\hat{k}}\bigr]  \delta Q^s_{\hat{k}}
   \\
  &\ +
  \bigl[\delta_{r,s}\, \lsb{QR}M^{r,r}_{\widehat{r+1},\hat{k}} 
             + \frac{1}{4}\delta^{(k-1)}_s\, \lsb{QR}M^{r,s}_{\hat{k}}\bigr]  \delta R^s_{\hat{k}}
 \Bigr] = \Lambda\, \delta Q^r_{\hat{k}}
 \nonumber
 \\
\label{eq:LAR}
 &\sum_{s=0}^{\mathcal{R}+1} \Bigl[
  \bigl[\delta_{r,s}\,\lsb{RQ}M^{r,r}_{\widehat{r+1},\hat{k}} 
             + \frac{1}{4}\delta^{(k-1)}_s\, \lsb{RQ}M^{r,s}_{\hat{k}}\bigr]  \delta Q^s_{\hat{k}}
  \\
  &\ +
  \bigl[\delta_{r,s}\, \lsb{RR}M^{r,r}_{\widehat{r+1},\hat{k}} 
             + \frac{1}{4}\delta^{(k-1)}_s\, \lsb{RR}M^{r,s}_{\hat{k}}\bigr]  \delta R^s_{\hat{k}}
 \Bigr] = \Lambda\, \delta R^r_{\hat{k}}
 \nonumber
 \end{align}
for $r = 0, \ldots, \mathcal{R}$ and
\begin{align}
\label{eq:LARb}
 \sum_{s=0}^{\mathcal{R}+1} 
     \frac{1}{4}\delta^{(k-1)}_s\,
 \Bigl[
             &\lsb{RQ}M^{\mathcal{R}+1,s}_{\hat{k}} \delta Q^s_{\hat{k}}
             \\
  &\quad+
              \lsb{RR}M^{\mathcal{R}+1,s}_{\hat{k}}  \delta R^s_{\hat{k}}
 \Bigr] 
  = 2\, \Lambda\, \delta R^{\mathcal{R}+1}_{\hat{k}}
 \nonumber
\end{align}
with $\delta Q^{\mathcal{R}+1}_{\hat{k}} = 0$ for $r=\mathcal{R}+1$, where
\begin{equation}
 \delta^{(k)}_s = p^{(k)}_s - p^{(k)}_{s+1}, \qquad
   p^{(k)}_s = \left\{\begin{array}{ll}
                p_s & s \leq k, \\
                \phantom{=} & \phantom{=} \\
                2 p_s & s > k.
                \end{array} \right.     
\end{equation}
The factor $2$ in \eqref{eq:LARb} follows from the 
multiplicity of the diagonal terms.
The coefficients of these equations are 
\begin{align}
 \lsb{QQ}M^{r,s}_{\hat{k}} &= \lsb{A}M^{r,s}_{\hat{k}} + {_{B}}M^{r,s}_{\hat{k}}, 
						    \\
 \lsb{RR}M^{r,s}_{\hat{k}} &= \lsb{A}M^{r,s}_{\hat{k}} + {_{B}}M^{r,s}_{\hat{k}}
 \\
             &\qquad\qquad- 2 \Lambda'(\overline{R}, 1)\, \delta_{r,s}\, \delta_{r,\mathcal{R}+1},
             \nonumber
						    \\
 \lsb{QR}M^{r,s}_{\hat{k}} &= \lsb{A}M^{r,s}_{\hat{k}} - {_{B}}M^{r,s}_{\hat{k}}, 
						    \\
 \lsb{RQ}M^{r,s}_{\hat{k}} &= \lsb{QR}M^{r,s}_{\hat{k}},
\end{align}
where ($C = A, B$)
\begin{align}
\nonumber 
  \lsb{C}M^{r,s}_{\hat{k}} &= 4 \frac{(C^{-1})_t}{C_{\hat{k}}}, 
             \quad\qquad\qquad k > t =\min(r,s),
             \\
\label{eq:Mrsk}
  \lsb{C}M^{r,s}_{\hat{k}} &= 2\sum_{u=k}^t p_u\bigl[ (C^{-1}_u)^2 - (C^{-1}_{u-1})^2\bigr] 
  \\
                 & \qquad + 
                 4 \frac{(C^{-1})_t}{C_{\widehat{t+1}}},
                 \quad\qquad
             k \leq t =\min(r,s),
             \nonumber
\end{align}
 and
 \begin{equation}
 \label{eq:Cm1}
  (C^{-1})_t =  \sum_{k=0}^{t} \frac{1}{p_k} \left[ 
              \frac{1}{C_{\hat{k}}} - \frac{1}{C_{\widehat{k+1}}} 
              \right].
 \end{equation}
Each eigenvalue has multiplicity
\begin{equation}
 \mu(k) = p_0\,\overline{\delta}_k, 
\end{equation}
with $\overline{\delta}_k$ defined in \eqref{eq:delta_bar} if $k>0$ and 
$\overline{\delta}_0 = 1/p_0$ for $k=0$.

If  $Q_0=R_0=0$ then $(A^{-1})_0 = (B^{-1})_0=0$ and
\begin{equation}
\forall k: \quad
  \lsb{C}M^{r,s}_{\hat{k}} = 0, \quad \text{if}\ \min(r,s)=0.
\end{equation}
The LA eigenvalue equations \eqref{eq:LAQ}-\eqref{eq:LAR}  for $r=0$ then decouple 
from the others and read:
\begin{eqnarray}
 \lsb{QQ}M^{0,0}_{\hat{1},\hat{k}}\, \delta Q^{0}_{\hat{k}} +
 \lsb{QR}M^{0,0}_{\hat{1},\hat{k}}\, \delta R^{0}_{\hat{k}} 
	  &=& \Lambda\,  \delta Q^{0}_{\hat{k}} , 
 \nonumber\\
 \lsb{RQ}M^{0,0}_{\hat{1},\hat{k}}\, \delta Q^{0}_{\hat{k}} +
  \lsb{RR}M^{0,0}_{\hat{1},\hat{k}}\, \delta R^{0}_{\hat{k}} 
	  &=& \Lambda\,  \delta R^{0}_{\hat{k}} .
 \end{eqnarray}
Comparison with  \eqref{eq:Rep_Q}-\eqref{eq:Rep_R} readily shows:
\begin{equation}
\label{eq:C34}
 \lsb{1,2}\Lambda(k) = \lsb{1,2}\Lambda(0;1,l), \qquad l = \max(1,k).
\end{equation}

\section{RS Saddle Point Hessian Eigenvalues}
\label{app:RS_Fluct}

For $\mathcal{R}=0$ and $Q_0=R_0=0$ the only non-null coefficients of the eigenvalues 
equations are
\begin{eqnarray}
 \lsb{QQ}M^{0,0}_{\hat{1},\hat{1}} &=& -\Lambda'(0, 0)
 			+ 2 \frac{1+\overline{R}^2}{\bigl(1-\overline{R}^2\bigr)^2} 
						    \\
 \lsb{RR}M^{0,0}_{\hat{1},\hat{1}} &=& -\Lambda'(0, 0) 
 			+ 2 \frac{1+\overline{R}^2}{\bigl(1-\overline{R}^2\bigr)^2} 
						    \\
 \lsb{RQ}M^{0,0}_{\hat{1},\hat{1}} &=& 
 			- 4 \frac{\overline{R}}{\bigl(1-\overline{R}^2\bigr)^2} 
						    \\
 \lsb{QR}M^{0,0}_{\hat{1},\hat{1}} &=& \lsb{RQ}M^{0,0}_{\hat{1},\hat{1}}.
\end{eqnarray}
and
\begin{equation}
 \lsb{RR}M^{1,1}_{\hat{0}}  = \lsb{RR}M^{1,1}_{\hat{1}} = -2\Lambda'(\overline{R},1) + 
                4\frac{1+\overline{R}^2}{\bigl(1-\overline{R}^2\bigr)^2}
\end{equation}
Then the eigenvalues in the Replicon sector are:
\begin{eqnarray}
 \lsb{1}\Lambda(0;1,1) &=& -\Lambda'(0,0) + \frac{2}{\bigl(1+\overline{R}\bigr)^2} 
 \\
 \lsb{2}\Lambda(0;1,1) &=& -\Lambda'(0,0) + \frac{2}{\bigl(1-\overline{R}\bigr)^2} 
\end{eqnarray}
each with multiplicity $\mu(0;1,1) = p_0(p_0-3)/2$.
In the LA sector the eigenvalues read:
\begin{eqnarray}
  \lsb{1}\Lambda(k)  &=& -\Lambda'(0,0) + \frac{2}{\bigl(1+\overline{R}\bigr)^2} 
 \\
 \lsb{2}\Lambda(k) &=& -\Lambda'(0,0) + \frac{2}{\bigl(1-\overline{R}\bigr)^2} 
\\
 \lsb{3}\Lambda(k) &=& -\frac{1}{2}\Lambda'(\overline{R},1) 
         + \frac{1+\overline{R}^2}{\bigl(1-\overline{R}^2\bigr)^2} 
\end{eqnarray}
with the $k=0$ eigenvalue (Longitudinal) of multiplicity $\mu(0) = 1$ and the $k=1$ eigenvalues 
(Anomalous) of multiplicity $\mu(1) = p_0-1$.

The relevant eigenvalues  for the stability of the RS solution with $\overline{R}\geq 0$ 
are, cf.  \eqref{eq:Lambda} for 
$\Lambda(x,y)$,\footnote{
If $\overline{R}\leq 0$ the eigenvalue $\protect\lsb{1}\Lambda(0;1,1)$  
is replaced by $\protect\lsb{2}\Lambda(0;1,1)$.
}
\begin{align}
	\lsb{1}\Lambda(0;1,1) &= -2\xi_2 + \frac{2}{(1+\overline{R})^2}\\
	\lsb{3}\Lambda(0) &= -(\xi_2 + 2\xi_4 + 3\xi_4\overline{R}^2) 
	   + \frac{1+\overline{R}^2}{\bigl(1-\overline{R}^2\bigr)^2}.
\end{align}
If $\overline{R}=0$ the Replicon eigenvalue is positive below the line $\xi_2 =1$
but  the LA eigenvalue is positive only below the  line $\xi_2+2\xi_4 = 1$. 
Above this line  an  the RS phase with $\overline{R}=0$ (IW phase) is replaced by the RS phase
with  $\overline{R}\not=0$ (Phase-Locking Wave phase).
The latter remain stable below the critical line
\begin{equation}
 \left\{ \begin{array}{lc}
  \displaystyle 
   \xi_2 = \frac{1}{(1+ t)^2} & \phantom{=}\\
   \phantom{=} & 0\leq t =|\overline{R}| \leq 1 \\
   \displaystyle
   \xi_4 = \frac{2 t}{(2+t^2) (1+t) (1- t^2)}.
 \end{array}
 \right.
\end{equation}
where the eigenvalue $\lsb{1}\Lambda(0;1,1)$ vanishes,
solid blue line in Figure \ref{fig:dynamic_lines}. 

\section{1RSB Saddle Point Hessian Eigenvalues}
\label{app:1RSB_Fluct}

We discuss only the solution $R_{\a\b} = Q_{\a\b}$, the analysis for the solution 
$R_{\a\b} = -Q_{\a\b}$ being similar.

The coefficients of the eigenvalue equation in the Replicon sector are
\begin{eqnarray}
 \lsb{QQ}M^{r,r}_{\hat{k},\hat{l}} &=& -\Lambda'(Q_r, Q_r) 
                                                            + \frac{1}{A_{\hat{k}}\,A_{\hat{l}}}
						    + \frac{1}{(B_{\hat{2}})^2}
						    \\
 \lsb{RR}M^{r,r}_{\hat{k},\hat{l}} &=& \lsb{QQ}M^{r,r}_{\hat{k},\hat{l}}
						    \\
 \lsb{RQ}M^{r,r}_{\hat{k},\hat{l}} &=& -\dot{\Lambda}(Q_r, Q_r) 
                                                            + \frac{1}{A_{\hat{k}}\,A_{\hat{l}}}
						    - \frac{1}{(B_{\hat{2}})^2}
						    \\
 \lsb{QR}M^{r,r}_{\hat{k},\hat{l}} &=& \lsb{RQ}M^{r,r}_{\hat{k},\hat{l}}.
\end{eqnarray}
with eigenvalues 
\begin{align}
	\lsb{1}\Lambda(r;k,l) &= 
	-\Lambda'(Q_r,Q_r) - \dot{\Lambda}(Q_r,Q_r) 
	            +\frac{2}{A_{\hat{k}}\,A_{\hat{l}}},
\\
	\lsb{2}\Lambda(r;k,l) &= 
		   -\Lambda'(Q_r,Q_r) + \dot{\Lambda}(Q_r,Q_r) 
	            +\frac{2}{(B_{\hat{2}})^2},     
\end{align}
with $r=0,1$ and $k,l \geq r+1$. 
Then for $(r,k,l) = (1,2,2)$:
\begin{align}
\label{eq:E7}
	\lsb{1}\Lambda(1;2,2) &= - 2(\xi_2 + 9\xi_4 Q_1^2)  
	            +\frac{2}{\bigl(1+\overline{R} - 2Q_1\bigr)^2},
\\
	\lsb{2}\Lambda(1;2,2) &= -2(\xi_2 + \xi_4 Q_1^2) 
	            +\frac{2}{\bigl(1-\overline{R}\bigr)^2}.
\end{align}
and multiplicity $\mu(1;2,2) = (p_0/2)(p_1-3)$.
For  $(r,k,l) = (0,2,2)$:
\begin{align}
	\lsb{1}\Lambda(0;2,2) &= - 2\xi_2  
	            +\frac{2}{\bigl(1+\overline{R} - 2Q_1\bigr)^2},
\\
	\lsb{2}\Lambda(0;2,2) &= - 2\xi_2
	            +\frac{2}{(1-\overline{R})^2},
\end{align}
and multiplicity $\mu(0;2,2) = (p_0/2)(1-1/p_1)^2(p_0 -p_1)$.
For $(r,k,l) = (0,1,1)$:
\begin{align}
	\lsb{1}\Lambda(0;1,1) &= - 2\xi_2  
	            +\frac{2}{\bigl[1+\overline{R} - 2(1-p_1) Q_1\bigr]^2}
\\
	{_2}\Lambda(0;1,1) &= \lsb{2}\Lambda(0;2,2)
\end{align}
and multiplicity $\mu(0;1,1) = (p_0/2)(1/p_1^2)(p_0 - 3p_1)$.
Finally for  $(r,k,l) = (0,2,1)$ or $(r,k,l) = (0,1,2)$:
\begin{align}
	\lsb{1}\Lambda(0;2,1) &= - 2\xi_2  
	\\
	            &+\frac{2}{\bigl(1+\overline{R} - 2Q_1\bigr)  \bigl[1+\overline{R} - 2(1-p_1) Q_1\bigr]} 
	            \nonumber
\\
	\lsb{2}\Lambda(0;2,1) &= \lsb{2}\Lambda(0;2,2)
\end{align}
and $\lsb{1,2}\Lambda(0;1,2) = \lsb{1,2}\Lambda(0;2,1)$. The multiplicity of these each of these 
eigenvalues is $\mu(0;2,1) = \mu(0;1,2) = (p_0/2)(1/ p_1)(1-1/p_1)(p_0 -2p_1)$.
The total dimension of the Replicon sector is then $p_0(p_0 - 5)$.

The calculation of the eigenvalues in the LA sector for $\mathcal{R}=1$ requires 
 the evaluation for each $k$ of the eigenvalues of a $5\times 5$ matrix.
Each eigenvalue has multiplicity respectively: $\mu(0) = 1$, 
$\mu(1) = p_0/p_1 -1$ and $\mu(2) = p_0 - p_0/p_1$.
The total dimension of the LA sector is then $5p_0$, which added to the
 dimension of the Replicon sector gives the total dimension $p_0^2$ of the
 $(Q_{\a\b},R_{\a\b})$ space.

For the solution with  $Q_0=R_0 = 0$ the $5\times 5$ matrix is partially diagonal and 
for each $k=0,1,2$ the first two 
eigenvalues  read, cf.  \eqref{eq:C34}: 
\begin{equation}
 \lsb{1,2}\Lambda(k) = \lsb{1,2}\Lambda(0;1,l), \quad l = \max(1,k).
\end{equation}

When $R_{\a\b}=Q_{\a\b}$, and $Q_0=R_0=0$, a straightforward calculation shows that
if  $\min(r,s) = 1$ then
 \begin{align}
 \nonumber
  \lsb{A}M^{r,s}_{\hat{2}} &= 4\frac{A^{-1}_1}{ A_{\hat{2}} } &\phantom{x}\\
  \lsb{A}M^{r,s}_{\hat{k}} &= 4\frac{A^{-1}_1}{ A_{\hat{2}} }  + 2p_1 (A^{-1}_1)^2, 
	&k = 0,1.
	\\
   \lsb{B}M^{r,s}_{\hat{k}} &= 0 &k = 0,1,2
	\nonumber
 \end{align}
while
 \begin{align}
  \lsb{A}M^{2,2}_{\hat{2}} &=
   2\left[\frac{1}{ (A_{\hat{2}})^2} + 2\frac{A^{-1}_1}{ A_{\hat{2}} } \right],
\\
  \lsb{A}M^{2,2}_{\hat{k}} &=
     2\left[\frac{1}{ (A_{\hat{2}})^2} + 2\frac{A^{-1}_1}{ A_{\hat{2}} }  + p_1(A^{-1}_1)^2 \right],
	\qquad k = 0,1
	\\
  \lsb{B}M^{2,2}_{\hat{k}}  &= \frac{2}{ (B_{\hat{2}})^2}, \qquad k = 0, 1, 2
  \nonumber
\end{align}
and
\begin{equation}
  \lsb{RR}M^{2,2}_{\hat{k}} = \lsb{A}M^{2,2}_{\hat{k}} + \frac{2}{ (B_{\hat{2}})^2} 
          - 2\Lambda'(\overline{R},1).
\end{equation}

The remaining three eigenvalues are then obtained from 
 an eigenvalue equation of the form
\begin{equation}
\label{eq:LAeigQ1R1}
  \left\{
    \begin{array}{ccc}
      \displaystyle
      (a+b) x + (c+b) y + d z &=& \Lambda\, x \\
      \displaystyle
      (c+b) x + (a+b) y + d z &=& \Lambda\, y \\
      \displaystyle
      \frac{b}{2}x + \frac{b}{2} y + \frac{e}{2} z &=& \Lambda\, z
    \end{array}
  \right.
\end{equation}
where
\begin{eqnarray}
 a &=& \lsb{QQ}M^{1,1}_{\hat{2},\hat{k}} = \lsb{RR}M^{1,1}_{\hat{2},\hat{k}}
 \nonumber\\
 b &=& \frac{1}{4}\delta^{(k-1)}_1\, \lsb{A}M^{1,1}_{\hat{k}} 
             = \frac{1}{4}\delta^{(k-1)}_1\, \lsb{A}M^{2,1}_{\hat{k}}
 \nonumber\\
c &=&  \lsb{QR}M^{1,1}_{\hat{2},\hat{k}} = \lsb{RQ}M^{1,1}_{\hat{2},\hat{k}} 
\\
d &=& \frac{1}{4}\delta^{(k-1)}_2\, \lsb{A}M^{1,2}_{\hat{k}}
\nonumber\\
e &=& \frac{1}{4}\delta^{(k-1)}_2\, \lsb{RR}M^{2,2}_{\hat{k}}.
\nonumber
\end{eqnarray}
The equation (\ref{eq:LAeigQ1R1}) admits two types of solution. The first  for
$y=-x$ and $z=0$ and leads to  the eigenvalue $\Lambda = a - c$,  i.e., 
\begin{equation}
 \lsb{3}\Lambda(k) = {_2}\Lambda(1;2;2), \qquad
                              k = 0, 1, 2.
\end{equation}
The other two  eigenvalues, for $x=y$ and $z\not=0$,  read instead
\begin{eqnarray}
 _{4}\Lambda(k) &=& \frac{1}{2} \left[T + \sqrt{T^2 - 4 \Delta}\right]
 \\
 _{5}\Lambda(k) &=& \frac{1}{2} \left[T - \sqrt{T ^2- 4 \Delta}\right]
\end{eqnarray}
where
\begin{equation}
  T = a + 2b + c + \frac{e}{2}, \qquad \Delta = \frac{e}{2}(a+2b+c) - db.
\end{equation}
The explicit form of $a, b, c, d,e$ depends on the value of $k$.
For $k=0,1$ one has:,
\begin{align}
  b &= (p_1-1) \left[ 2\frac{A^{-1}_1}{ A_{\hat{2}} }  + p_1 (A^{-1}_1)^2\right],\\
  d &=  2\frac{A^{-1}_1}{ A_{\hat{2}} }  + p_1 (A^{-1}_1)^2,                \\
 e & =
     -\Lambda'(\overline{R},1)  + \frac{1}{(A_{\hat{2}})^2} + 2\frac{A^{-1}_1}{ A_{\hat{2}} }  
     \nonumber
     \\
       &\phantom{=======}+ p_1 (A^{-1}_1)^2 + \frac{1}{(B_{\hat{2}})^2},
\end{align}
 and  
\begin{equation}
 \begin{split}
a+&2b+c =   -\Lambda'(Q_1,Q_1) -\dot{\Lambda}(Q_1,Q_1) 
\\
&+\frac{2}{(A_{\hat{2}})^2} +
    2(p_1-1)\left[2\frac{A^{-1}_1}{ A_{\hat{2}} }  + p_1 (A^{-1}_1)^2\right].
\end{split}
\end{equation}

For  $k=2$ one has instead:
\begin{align}
  b &= (p_1-2) \frac{A^{-1}_1}{ A_{\hat{2}} },\\
  d &=  2\frac{A^{-1}_1}{ A_{\hat{2}} },                \\
 e & =
     -\Lambda'(\overline{R},1)  + \frac{1}{(A_{\hat{2}})^2} + 2\frac{A^{-1}_1}{ A_{\hat{2}} }  
       + \frac{1}{(B_{\hat{2}})^2},
\end{align}
 and  
\begin{equation}
 \begin{split}
a+2b+c =   -\Lambda'(&Q_1,Q_1) -\dot{\Lambda}(Q_1,Q_1) 
\\
&+\frac{2}{(A_{\hat{2}})^2} +
    2(p_1-2)\frac{A^{-1}_1}{ A_{\hat{2}} }.
\end{split}
\end{equation}
The eigenvalues $\lsb{4,5}\Lambda(k)$  are always positive or complex, with positive real part 
in the whole 1RSB phase with $R_1=Q_1$.

The relevant eigenvalues for the stability of the 1RSB solution 
are ${_1}\Lambda(1;2,2)$, which controls the stability with respect to $Q_1$, 
and  ${_1}\Lambda(0;1,1)$, which controls the fluctuations with
respect to $Q_0=0$.  The vanishing of ${_1}\Lambda(1;2,2)$ leads to the {\sl marginal condition},
while the vanishing of ${_1}\Lambda(0;1,1)$ marks the end of the 1RSB phase and the 
appearance of a 1FRSB phase. 
Using the saddle point equation, the critical 1RSB-1FRSB line in the limit $p_0=n\to 0$ is 
\begin{equation}
 \left\{ \begin{array}{lc}
  \displaystyle 
   \xi_2 = \frac{1}{\overline{a}^2 x^2}(1 - y_0 + xy_0)^2 & \phantom{=}\\
   \phantom{=} & \phantom{=} \\
   \displaystyle
   \xi_4 = \frac{4}{3\overline{a}^2}\frac{(1-y_0 + xy_0)^4}{x^2 y (1-y_0)}
 \end{array}
 \right.
\end{equation}
where $x=p_1 \in[0,1]$ and $y_0=0.38957\ldots$ is the solution of $z(y_0) = (1+y_0)/2$
with $z(y)$ defined in \eqref{eq:def_zFunction}, solid black line in Figure \ref{fig:dynamic_lines}.
Note that for $x=1$ 
\begin{equation}
  	{_1}\Lambda(0;1,1) = - 2\xi_2  
	            +\frac{2}{\bigl(1+\overline{R}\bigr)^2}
\end{equation}
and one recovers the relevant Replicon eigenvalue of the RS solution. As a consequence the 
RS and 1RSB critical lines meet at the tricritical point. This is a general result.



%

\end{document}